\documentclass[reqno,a4paper,11pt]{amsart}
\usepackage{amsmath, graphicx}
\usepackage{amstext,amsfonts,amsbsy,eucal,amssymb}

\numberwithin{equation}{section}

\newtheorem{theorem}{Theorem}[section]
\newtheorem{lemma}[theorem]{Lemma}
\newtheorem{proposition}[theorem]{Proposition}
\newtheorem{corollary}[theorem]{Corollary}

\theoremstyle{definition}
\newtheorem{definition}[theorem]{Definition}
\newtheorem{example}[theorem]{Example}
\newtheorem{remark}[theorem]{Remark}

\newcommand{\e}{\operatorname{\mathrm{e}}}

\newcommand{\R}{\operatorname{\mathbb{R}}}
\newcommand{\Z}{\operatorname{\mathbb{Z}}}
\newcommand{\C}{\operatorname{\mathbb{C}}}
\newcommand{\ve}{\operatorname{\varepsilon}}
\newcommand{\Div}{\operatorname{\mathrm{Div}}}
\newcommand{\Log}{\operatorname{\mathrm{Log}}}
\newcommand{\I}{\operatorname{\sqrt{-1}}}

\newcommand{\cS}{\operatorname{\mathcal{S}}}

\def\<{\operatorname{\langle}}
\def\>{\operatorname{\rangle}}
\def\pic{{\rm Pic}}

\def\ve{\varepsilon}

\makeatletter
\def\iddots{\mathinner{\mkern1mu\raise\p@
    \hbox{.}\mkern2mu\raise4\p@\hbox{.}\mkern2mu
    \raise7\p@\vbox{\kern7\p@\hbox{.}}\mkern1mu}}
\def\adots{\mathinner{\mkern2mu\raise\p@\hbox{.} %% yhmath.sty'©'ç
 \mkern2mu\raise4\p@\hbox{.}\mkern1mu
 \raise7\p@\vbox{\kern7\p@\hbox{.}}\mkern1mu}}
\makeatother

\begin{document}

\baselineskip 16pt

\title{\large 
A tropical analogue of Fay's trisecant identity and
the ultra-discrete periodic Toda lattice
}

\author{Rei Inoue}
\address{Faculty of Pharmaceutical Sciences,
Suzuka University of Medical Science, \\
3500-3 Minami-tamagaki, Suzuka, Mie, 513-8670, Japan}
\email{reiiy@suzuka-u.ac.jp}

\author{Tomoyuki Takenawa} 
\address{Faculty of Marine Technology, 
Tokyo University of Marine Science and Technology,  
2-1-6 Etchu-jima, Koto-ku, Tokyo, 135-8533, Japan}
\email{takenawa@kaiyodai.ac.jp}

\begin{abstract}
We introduce a tropical analogue of Fay's trisecant identity 
for a special family of hyperelliptic tropical curves. 
We apply it to obtain the general solution of 
the ultra-discrete Toda lattice with periodic boundary conditions 
in terms of the tropical Riemann's theta function.    
\end{abstract}

\keywords{tropical geometry, Riemann's theta function, 
Toda lattice, spectral curve}

%\renewcommand{\subjclassname}{%
%  \textup{2000} Mathematics Subject Classification}
%\subjclass{Primary: 37J35. Secondary: 14H70, 14H40.}

\maketitle

%%%%%%%%%%%%%%%%%%%%%%
\section{Introduction}
%%%%%%%%%%%%%%%%%%%%%%%%%%%%%%%%%%%%%%%%
\subsection{Background and main results}
%%%%%%%%%%%%%%%%%%%%%%%%%%%%%%%%%%%%%%%%

Fay's trisecant identity is an important and 
special identity satisfied by 
Riemann's theta functions for the Jacobian of a curve
\cite{Fay73,Mumford-Book},
and plays a crucial role in studying classical 
integrable systems. 
For instance, 
Fay's trisecant identity gives a solution to the Hirota-Miwa equation
~\cite{Hirota81, Miwa82}
from which many soliton equations derive.

Recently a tropical analogue of Riemann's theta function was
introduced in some contexts 
(\cite{KS06,MikhaZhar06}, in \cite{Nobe05} the tropical Riemann's 
addition formula is introduced). 
The first aim of this paper is to introduce a tropical version of 
Fay's trisecant identity (Theorem \ref{tropicalFay}).
Due to a technical difficulty, our result is restricted to 
the case of some special hyperelliptic tropical curves,
but we expect that it also holds for general tropical curves.

On the other hand, in \cite{InoueTakenawa07} 
we studied the ultra-discrete periodic Toda lattice (UD-pToda)
and proposed a method to deal with integrable cellular 
automata via the tropical algebraic geometry.
Since the UD-pToda is closely related 
to an important soliton cellular automata called the box and ball system
\cite{TakahashiSastuma90},  
it is regarded as a fundamental
object in the studies of integrable automata.
The UD-pToda is defined by the following piecewise-linear map
\begin{align}\label{UD-pToda}
  \begin{split}
  &Q_n^{t+1} = \min[W_n^t, Q_n^t-X_n^t],
  \qquad 
  W_n^{t+1} = Q_{n+1}^t+W_n^t - Q_n^{t+1},
  \\
  &\text{ with }
  X_n^t = \min_{k=0,\cdots,g}\bigl[\sum_{l=1}^k (W_{n-l}^t - Q_{n-l}^t)\bigr],
  \end{split}
\end{align}
on the phase space: 
\begin{align}\label{phasespace-T}
\mathcal{T} = \{(Q,W) = 
(Q_1,\cdots,Q_{g+1},W_1,\cdots,W_{g+1}) 
\in \R^{2(g+1)} ~|~ \sum_{n=1}^{g+1} Q_n < \sum_{n=1}^{g+1} W_n \}.
\end{align}
Here we fix $g \in \Z_{>0}$ and 
assume periodicity $Q_{n+g+1}^t = Q_n^t$ and $W_{n+g+1}^t = W_n^t$.
This system has a tropical spectral curve $\Gamma$, and
we conjectured that its general isolevel set is 
isomorphic to the tropical Jacobian $J(\Gamma)$
of $\Gamma$ (the cases of $g=1,2,3$ were proved).  
It is expected that the solution for the UD-pToda is written 
in terms of the tropical Riemann's theta function associated with $\Gamma$ 
as the classical cases,
which is another aim of this paper. 
For this purpose we transform \eqref{UD-pToda} into an equation for 
the quasi-periodic function $T_n^t$,
\begin{align}\label{T-eq}
  T_n^{t+2}+T_n^t
  = 
  \min_{k=0,\cdots,g}\bigl[k L + 2 T_{n-k}^{t+1} + T_{n+1}^t+T_{n}^t
          - (T_{n-k+1}^t + T_{n-k}^t)\bigr],
\end{align}
via
\begin{align}\label{WQ-T}
  \begin{split}
  &W_n^t = L + T_{n-1}^{t+1} + T_{n+1}^{t} - T_{n}^{t} - T_{n}^{t+1}+C_g,
  \\
  &Q_n^t = T_{n-1}^{t} + T_{n}^{t+1} - T_{n-1}^{t+1} - T_{n}^{t}+C_g.
  \end{split}
\end{align}
Here $L$ and $C_g$ 
are determined by $\{Q_n^t,W_n^t\}_{n=1,\cdots,g+1}$
and preserved by the map \eqref{UD-pToda}.
We show that the equation \eqref{T-eq} is essentially equivalent to a
tropical bilinear equation (Proposition \ref{prop:tau-ptau}):
\begin{align}\label{UD-tau}
  T_{n}^{t-1} +  T_{n}^{t+1}
  = 
  \min[2 \, T_{n}^{t}, ~T_{n-1}^{t+1} + T_{n+1}^{t-1} +L],
\end{align}
which is turned out to be a particular case of
Fay's trisecant identity for the tropical Riemann's theta function
(Corollary \ref{prop:tau-theta}).
Finally we obtain the general solution for the UD-pToda 
in terms of the tropical Riemann's theta function
(Theorem \ref{theta-tau}).

In the following three subsections, we introduce some fundamental notions 
related to tropical geometry used in this article and
the background results on the relation between the 
discrete periodic Toda lattice (D-pToda) and the UD-pToda.

%%%%%%%%%%%%%%%%%%%%%%%%%%%%%%
\subsection{Tropical Jacobian}
%%%%%%%%%%%%%%%%%%%%%%%%%%%%%%

\begin{definition}\cite{Mikhalkin03}
A finite connected graph $\Sigma \hookrightarrow \R^2$ is called 
a (plane) tropical curve
if the weight $w_e \in \Z_{_>0}$ is defined for each edge $e$ and
the following is satisfied:\\
(i) The tangent vector 
of each edge is rational.\\
(ii) For each vertex $v$, let $e_1,\dots,e_n$ be the oriented edges 
outgoing from $v$.  
 Then the primitive tangent vectors $\xi_{e_k}$ of $e_k$  
  satisfy
$
  \sum_{k=1}^{n} w_{e_k} \xi_{e_k} = 0
$.

Further, a tropical curve $\Sigma$ is called smooth 
if the following is satisfied:
\\
(iii) All the weights are $1$.\\
(iv) Each vertex $v$ is $3$-valent
and the primitive tangent vectors satisfy $|\xi_{e_i}\wedge \xi_{e_j}|=1$ for 
$i\neq j \in \{1,2,3\}$.
\end{definition}

Let $\tilde{\Gamma}$ be a smooth tropical curve,
$\Gamma:= \tilde{\Gamma} \setminus \{\text{infinite edges}\}$ be
the maximal compact subset of $\tilde{\Gamma}$,
$g$ be the genus of $\Gamma$, i.e. $g= \mathrm{dim} H_1(\Gamma,\Z)$, and
$B_1,\dots, B_g$  be a basis of $H_1(\Gamma,\Z)$.

Following \cite[\S 3.3]{MikhaZhar06}, we equip $\Gamma$ with the structure of 
a metric graph. 
For points $x,y$ on some edge $e$ in $\Gamma$,
we define the weighted distance $d(x,y)$ by
$$
  d(x,y) = \frac{\parallel x-y \parallel}{\parallel \xi_e \parallel},
$$
where $\parallel . \parallel$ denotes the Euclidean norm in 
$\R^2$.
With this distance 
the tropical curve $\Gamma$ becomes a metric graph.
The metric on $\Gamma$ defines a symmetric bilinear form $\<\cdot , \cdot \>$ 
on the space of paths on $\Gamma$ as follows:
for a non-self-intersecting path $\rho$, set $\<\rho, \rho\> 
:= \mathrm{length}_d(\rho)$,
and extending it to any pairs of paths bilinearly.

\begin{definition}\cite[\S 6.1]{MikhaZhar06}
  The tropical Jacobian of $\Gamma$ is a $g$ dimensional real torus 
  defined as
  $$ 
    J(\Gamma) = H_1(\Gamma,\R) / H_1(\Gamma,\Z)
    \simeq \R^g / K \Z^g.
  $$
  Here $K \in M_g(\R)$ are given by
  \begin{align}\label{periodK}
    K_{ij} =& \<B_i, B_j\>.
  \end{align}
\end{definition}
Since $\<\cdot , \cdot \>$ is nondegenerate,
$K$ is symmetric and positive definite.
In particular, $J(\Gamma)$ is called principally polarized.

Fix $P_0 \in \Gamma$.
Let $\Div(\Gamma)$ be the set of divisors on $\Gamma$.
Following \cite{MikhaZhar06} we define the tropical Abel-Jacobi map 
$\eta :~ \Div(\Gamma) \to J(\Gamma)$ by 
\begin{align}
  \label{UD-AJ}
  \sum_i k_i P_i \mapsto \sum_i k_i \int_{P_0}^{P_i} :=
  \sum_i k_i{}(\<\Gamma_{P_0}^{P_i},B_1\>,
   \cdots,\<\Gamma_{P_0}^{P_i},B_g\>)
,
\end{align}
where $k_i$ is a nonzero integer and 
$\Gamma_{P_0}^{P_i}$ is a path from $P_0$ to $P_i$ on $\Gamma$.
For example, 
the divisor $P_2-P_1$ is mapped to 
$$\eta(P_2-P_1)=\int_{P_1}^{P_2} =
(\<\Gamma_{P_1}^{P_2},B_1\>,
   \cdots,\<\Gamma_{P_1}^{P_2},B_g\>).$$

%%%%%%%%%%%%%%%%%%%%%%%%%%%%%%%%
\subsection{Ultra-discrete limit}
%%%%%%%%%%%%%%%%%%%%%%%%%%%%%%%%

The ultra-discrete limit (UD-limit) 
links discrete dynamical systems to cellular automata,  
and algebraic curves to tropical curves.
This limit is also called tropicalization.
We define a map $\Log_\ve : \R_{>0} \to \R$ with an infinitesimal parameter 
$\ve > 0$ by
\begin{align}
  \label{loge-map}
  \Log_{\ve} : x \mapsto - \ve \log x.
\end{align}
For $x > 0$, we define $X \in \R$ by $x = \e^{-\frac{X}{\ve}}$.
Then the limit $\ve \to 0$ of $\Log_\ve (x)$ converges to $X$.    
The procedure $\lim_{\ve \to +0} \Log_\ve$ with the scale transformation as 
$x = \e^{-\frac{X}{\ve}}$
is called the UD-limit. 

We summarize this procedure in more general setting:
\begin{proposition}
For $A,B,C\in \R$ and $k_a,k_b,k_c>0$, 
set 
$$a=k_a e^{-\frac{A}{\ve}},\ b=k_b e^{-\frac{B}{\ve}}, 
\ c=k_c e^{-\frac{C}{\ve}}.$$
Then the UD-limit of the equations
$$({\rm i})\ a+b=c,\qquad ({\rm ii})\ ab=c,\qquad ({\rm iii})\ a-b=c$$
yields the followings:
\begin{align*}
  &({\rm i})\ \min [A,B]=C, \qquad ({\rm ii})\ A+B=C,
  \\
  &({\rm iii}) 
  \begin{cases} A=C& \ \ ({\rm if} \ \ A<B, 
  ~{\rm or }~ A=B~ {\rm and }~ k_a > k_b)\\
  {\rm contradiction}& \ \ ({\rm otherwise})
  \end{cases}.
\end{align*}
\end{proposition}

%%%%%%%%%%%%%%%%%%%%%%%%%%%%%%%%%%%%%%%%%%%%%%%%
\subsection{From D-pToda lattice to UD-pToda}
%%%%%%%%%%%%%%%%%%%%%%%%%%%%%%%%%%%%%%%%%%%%%%%%

We briefly review
the D-pToda and the way to obtain the UD-pToda.
Fix $g \in \Z_{>0}$.
The $(g+1)$-periodic Toda lattice of discrete time $t \in \Z$ 
\cite{HirotaTsujiImai} is given by the rational map 
on the phase space 
$\mathcal{U} = \{(I_1,\cdots,I_{g+1}, V_1,\cdots, V_{g+1})\} 
\simeq \C^{2(g+1)}$: 
\begin{align}
  \label{d-Toda}
  I_n^{t+1} = I_n^t+V_n^t - V_{n-1}^{t+1},
  \qquad
  V_n^{t+1} = \frac{I_{n+1}^t V_n^t}{I_n^{t+1}},
\end{align}
where we assume the periodicity $I_{n+g+1}^t = I_n^t$ and 
$V_{n+g+1}^t = V_n^t$.
This system has the $(g+1)$ by $(g+1)$ Lax matrix:
\begin{align}
  \label{d-Toda-Lax}
  L^t(y) = 
    \begin{pmatrix} 
      a_1^t & 1 & & & (-1)^g \frac{b_1^t}{y} \\
      b_2^t & a_2^t & 1 & & \\
      & \ddots & \ddots & \ddots & \\
      & & b_{g}^t & a_{g}^t & 1 \\
      (-1)^g y & & & b_{g+1}^t & a_{g+1}^t \\
    \end{pmatrix}.
\end{align} 
Here $y \in \C$ is a spectral parameter,
and we set $a_n^t = I_{n+1}^t+V_n^t$, $b_n^t = I_n^t V_n^t$.
The evolution \eqref{d-Toda} preserves 
the algebraic curve $\gamma$ given by 
$f(x,y)=y \det (x \mathbb{I}_{g+1}+L^t(y)) = 0$. 
When we fix a polynomial $f(x,y)$:
\begin{align}
  \label{complex-curve}
  f(x,y) = y^2+y (x^{g+1}+c_{g} x^{g}+\cdots+c_1 x+c_0)+c_{-1},
\end{align}
the isolevel set $\mathcal{U}_c$ for \eqref{d-Toda} is 
$$
  \mathcal{U}_c = \{ (I_n^t,V_n^t)_{n=1,\cdots,g+1} \in \mathcal{U} ~|~
  y \det (x \mathbb{I}_{g+1}+L^t(y)) = f(x,y) \}. 
$$
It is known that for generic $f(x,y)$,
$\mathcal{U}_c$ is isomorphic to an affine part of the Jacobian 
$\mathrm{Jac}(\gamma)$ of $\gamma$
\cite{MM79}.

\begin{figure}
\begin{center}
\unitlength=1.2mm
\begin{picture}(100,70)(0,10)
%\put(0,10){\line(1,0){20}}
%\put(20,10){\vector(1,0){80}}
%\put(100,7){$X$}
%\put(10,0){\line(0,1){10}}
%\put(10,10){\line(0,1){65}}
%\put(10,75){\vector(0,1){5}}
%\put(7,80){$Y$}
%\put(-10,5){$(0,(g+1)C_g)$}

%\put(25,13){$\lambda_1-\lambda_0$}
%\put(33,22){$\lambda_2-\lambda_1$}
%\put(58,33){$\lambda_g-\lambda_{g-1}$}

%\put(25,69){$\lambda_1-\lambda_0$}
%\put(32,62){$\lambda_2-\lambda_1$}
%\put(58,50){$\lambda_g-\lambda_{g-1}$}

\put(20,10){\line(-2,-3){6}}
\put(20,10){\line(4,5){8}}
\put(28,20){\line(1,1){8}}
\put(36,28){\line(2,1){10}}
\put(53,35){\line(5,1){15}}
\put(68,38){\line(1,0){15}}

\put(20,75){\line(-2,3){6}}
\put(20,75){\line(4,-5){8}}
\put(28,65){\line(1,-1){8}}
\put(36,57){\line(2,-1){10}}
\put(53,50){\line(5,-1){15}}
\put(68,47){\line(1,0){15}}

\put(20,10){\line(0,1){65}}
\put(28,20){\line(0,1){45}}
\put(36,28){\line(0,1){29}}
\put(56,35.5){\line(0,1){14}}
\put(68,38){\line(0,1){9}}

\put(48,42){$\cdots$}

%\put(18.5,43){$L$}
%\put(18.5,7){$C_g$}
%\put(27,43){$p_1$}
%\put(35,43){$p_2$}
%\put(43,43){$\cdots$}
%\put(51,43){$p_{g-1}$}
%\put(67,43){$p_{g}$}

\put(24,42){\oval(4,45)}
\put(26,30){\vector(0,1){3}}
\put(22.5,30){$B_1$}
\put(32,42){\oval(4,27)}
\put(34,35){\vector(0,1){3}}
\put(30.5,35){$B_2$}
\put(62,42){\oval(8,8)}
\put(66,41){\vector(0,1){3}}
\put(61.5,41){$B_g$}

\end{picture}
\caption{Tropical curve $\tilde{\Gamma}$}\label{Gamma}
\end{center}
\end{figure}
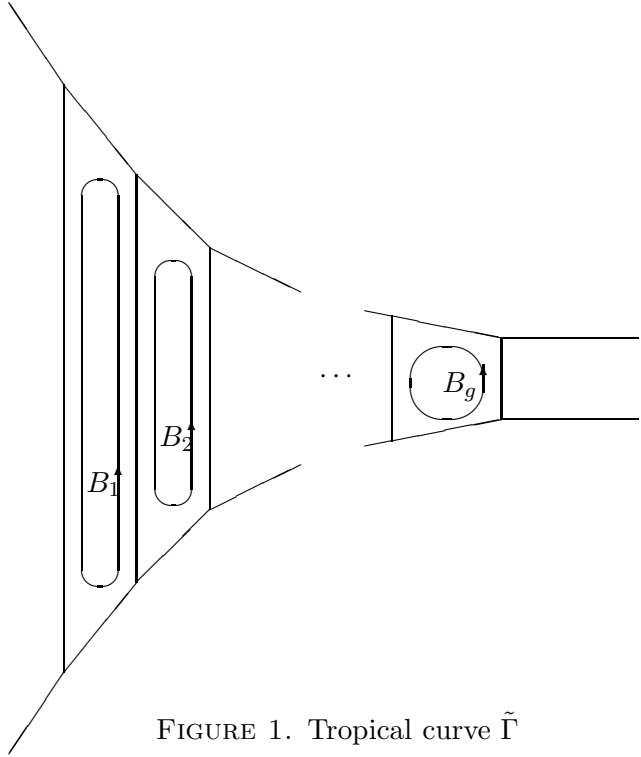

Eq. \eqref{d-Toda} are rewritten as \cite{KimijimaTokihiro02}
\begin{align}\label{d-Toda2}
  \begin{split}
  &I_n^{t+1} = V_n^t + 
              I_n^t \frac{1-\frac{\prod_{n=1}^{g+1} V_n^t}
                                 {\prod_{n=1}^{g+1} I_n^t}}
                         {1+\frac{V_{n-1}^t}{I_{n-1}^t}+\cdots+
                            \frac{V_{n-1}^t \cdots V_{n-g}^t}
                                 {I_{n-1}^t \cdots I_{n-g}^t}}, 
  \\
  &V_n^{t+1} = \frac{I_{n+1}^t V_n^t}{I_n^{t+1}}.
  \end{split}
\end{align}
Under the condition $\prod_{n=1}^{g+1} I_n^t > \prod_{n=1}^{g+1} V_n^t$,
the UD-limit of \eqref{d-Toda2} 
with the scale transformation
$I_n^t = \e^{-\frac{Q_n^t}{\ve}}, V_n^t = \e^{-\frac{W_n^t}{\ve}}$
gives the UD-pToda \eqref{UD-pToda}.
The UD-pToda preserves the tropical curve  
$\tilde{\Gamma} \subset \mathbb{R}^2$ given by \cite{InoueTakenawa07}:
\begin{align}\label{trop-curve}
  \begin{split}
  &\tilde{\Gamma}=\{(X,Y)\in \R^2 \ ~|~ F(X,Y)\mbox{ is not smooth}\}, 
  \\
  &\mbox{ ~where } F(X,Y)=
  \min[2Y, Y+\min[(g+1)X, gX+C_g, \cdots, X+C_1,C_0],C_{-1}].
  \end{split}
\end{align}
Here $C_i$'s are regarded as tropical polynomials on $\mathcal{T}$
\eqref{phasespace-T}.
For generic $(Q,W)\in \mathcal{T}$,
$C=(C_{-1},C_0,\cdots,C_g) \in \R^{g+2}$ satisfies 
\begin{align}
  \label{CD-condition}  
  C_{-1} > 2 C_0, ~  
  C_i+C_{i+2} > 2 C_{i+1} ~ (i=0,\cdots,g-2),~
  C_{g-1} > 2 C_g,
\end{align}
and we refer it as the generic condition.
With this condition, $\tilde{\Gamma}$ becomes a smooth tropical curve.
We show the shape of $\Tilde{\Gamma}$ 
with the basis $B_i$'s of $H_1(\Gamma,\Z)$
at Figure~\ref{Gamma}. %~\ref{Gamma-metric}

%%%%%%%%%%%%%%%%%%%%%%%%%%%%%%%%%%
\subsection{Content}
%%%%%%%%%%%%%%%%%%%%%%%%%%%%%%%%%%
In \S 2, we introduce a tropical analogue of 
Fay's trisecant identity (Theorem \ref{tropicalFay}) and obtain the bilinear
form of the UD-pToda \eqref{UD-tau} as its particular case.
In \S 3, we study the relation between the UD-pToda \eqref{UD-pToda}
and the bilinear form \eqref{UD-tau} (Lemma \ref{lemma:tau-QW}), and 
obtain the general solutions
in terms of the tropical Riemann's theta function (Theorem \ref{theta-tau}).
Appendix A is devoted to prove Theorem \ref{UDAJ}
which is a key to Theorem \ref{tropicalFay}.

%%%%%%%%%%%%%%%%%%%%%%%%%%%%%%%%%%
\subsection{Notations}
%%%%%%%%%%%%%%%%%%%%%%%%%%%%%%%%%%
We use the following notations of vectors in $\R^g$:
\begin{align*}
\vec{g}&=(g,g-1,\dots,1),\\
{\bf e}_i&: \mbox{the $i$-th vector of standard basis of } \R^g,\\
{\mathbb I}&=(1,1,\dots,1)={\bf e}_1+\cdots+{\bf e}_g,\\
{\mathbb I}_k&=(1,\dots,1,0,\dots,0)={\bf e}_1+\cdots+{\bf e}_k.
\end{align*}

%%%%%%%%%%%%%%%%%%%%%%%%%%%%%
\subsection*{Acknowledgement}
%%%%%%%%%%%%%%%%%%%%%%%%%%%%%

T.~T. appreciates the assistance from the Japan Society for the
Promotion of Science.
R.~I. is supported by the Japan Society for the
Promotion of Science, Grand-in-Aid for Young Scientists (B) (19740231).

%%%%%%%%%%%%%%%%%%%%%%%%%%%%%%%%%%%%%%
\section{Fay's trisecant identity and its tropicalization}
%%%%%%%%%%%%%%%%%%%%%%%%%%%%%%%%%%%%%%

\subsection{Fay's trisecant identity for hyperelliptic curves}
\label{faysidentity}

Let $\gamma$ be the hyperelliptic curve  
given by $v^2=\prod_{i=1}^{2g+2}(u-u_i)$,
which defines the two-sheeted covering 
$u_{\pm}$ of $u$ with branches 
$[u_{2k+1}, u_{2k+2}]$ $(k=0,1,2,\dots,g)$.
Choose the basis $a_1,\dots,a_g,b_1,\dots,b_g$ of $H_1(\gamma,\Z)$  
as usual as 
\begin{itemize}
  \item[(a)] $a_k$ goes the circuit around the branch 
  $[u_{2k+1},u_{2k+2}]$ on $u_+$ ,
  \\
  \item[(b)] $b_k$ goes on the upper half of $u_+$ from $[u_1,u_2]$
  to $[u_{2k+1},u_{2k+2}]$ and goes on the lower half
  of $u_-$ from $[u_{2k+1},u_{2k+2}]$ to $[u_1,u_2]$.
\end{itemize}
Let $\omega_1,\dots,\omega_g$ be a basis of 
the holomorphic differentials $H^0(\gamma,\Omega^1)$
normalized so that the period matrix with respect to
$a_1,\dots,a_g,b_1,\dots,b_g$ has the form $(I,\Omega)$,
where $I$ is the $g \times g$ identity matrix and $\Omega \in M_g(\C)$
is the symmetric matrix whose imaginary part is positive definite.
We write $\mathrm{Jac}(\gamma) = \C^g/(\Z^g I + \Z^g \Omega)$
for the Jacobian of $\gamma$.

We define a generalization of Riemann's theta function as
$$\theta[\alpha,\beta]({\bf z}) := 
\exp \{\pi \I(\beta \Omega \beta^{\bot}+ 2\beta ({\bf z}+\alpha)^{\bot})\} 
\theta({\bf z}+\Omega\beta+\alpha)$$
for $\alpha, \beta \in \R^g$ and ${\bf z}\in \C^g$,
where Riemann's theta function $\theta({\bf z})$ is 
\begin{align}\label{R-theta}
  \theta({\bf z})= \theta[0,0]({\bf z}) 
  = \sum_{{\bf m}\in \Z^g} \exp \{\pi \I 
({\bf m}\Omega {\bf m}^{\bot}+2{\bf m}{\bf z}^{\bot})\}.
\end{align}

We set $Q_i=(u_i,0) \in \gamma$, 
and take $Q_1$ as a base point of the Abel-Jacobi map, 
\begin{align}\label{A-J}
  \eta : \mathrm{Div}(\gamma) \to \mathrm{Jac}(\gamma) 
  ; ~ \sum_i k_i P_i \mapsto 
  \sum_i k_i (\int_{Q_1}^{P_i} \omega_j)_{j=1,\dots,g},
\end{align}
where $k_i \in \Z, ~ P_i \in \gamma$.
Let $K_\gamma$ be the canonical divisor on $\gamma$
and $\phi$ be the hyperelliptic involution of $\gamma$ 
(interchanging the two sheets $u_\pm$). 
One sees that
$\Delta :=-Q_1+Q_3+Q_5+\cdots+Q_{2g+1} \in \pic^{g-1}(\gamma)$
is a theta characteristic
(i.e. $2 \Delta = K_\gamma$ in $\pic^{2g-2}(\gamma)$) and that
$D := P+\phi(P) \in \pic^2(\gamma)$ for $P\in \gamma$ satisfies
$\eta(D)=0$.  
Via the Abel-Jacobi map \eqref{A-J},
the theta characteristics correspond to the half-periods of 
$\mathrm{Jac}(\gamma)$. For instance we have
$$\eta(\Delta)=\frac12\left(
\begin{array}{cccc}g&g-1&\cdots&1\\[1mm] 1&1&\cdots&1\end{array}
\right)_{\Omega},$$
where $\begin{pmatrix} \alpha \\ \beta \end{pmatrix}_\Omega
= \alpha I + \beta \Omega \in \C^g$ for $\alpha, \beta \in \R^g$.

For $m = 0,1,\cdots, [\frac{g+1}{2}]$,
let $\{i_1,\cdots, i_{g+1-2m} \}$ be a subset of $\{1,2,\cdots,2g+2\}$.
The following is known \cite[pp 12-15]{Fay73}: 
for $m=0$, $\eta(\Delta+D-\sum_{k=1}^{g+1} Q_{i_k})$ 
are the non-singular even half-periods,
while for $m=1$, $\eta(\Delta-\sum_{k=1}^{g-1} Q_{i_k})$ are 
the non-singular odd half-periods, and
for $m>1$, $\eta(\Delta-(m-1)D-\sum_{k=1}^{g+1-2m} Q_{i_k})$ are 
the even (odd) singular half-periods of multiplicity $m$ when 
$m$ is even (odd).
By using the formulae: 
\begin{align*}
&\eta(Q_2)=\frac12\left(\begin{array}{cccc}
1&1&\cdots&1\\0&0&\cdots&0
\end{array}\right)_{\Omega}, \quad
\eta(Q_{2k+1})=
\frac12\left(\begin{array}{ccccccc}
0&\cdots&0&\begin{array}[b]{c}\text{\scriptsize{$k$}}\\ 
\text{\scriptsize{$\vee$}}\\1\end{array}&1&\cdots&1\\
0&\cdots&0&1&0&\cdots&0\end{array}\right)_{\Omega}\\
&\eta(Q_{2k+2})=
\frac12
\left(\begin{array}{ccccccc}
0&\cdots&0&\begin{array}[b]{c}\text{\scriptsize{$k$}}\\ 
\text{\scriptsize{$\vee$}}\\0\end{array}&1&\cdots&1\\
0&\cdots&0&1&0&\cdots&0
\end{array}\right)_{\Omega}
\end{align*}
we obtain the following:

\begin{proposition}\label{nonsing.odd}
For a hyperelliptic curve $\gamma$ and a half integer vector
$\beta(\neq 0) \in \frac12 \Z^g/\Z^g$, set $\alpha\in\frac12 \Z^g$ as 
$$\alpha=(0,\cdots,0,\begin{array}[b]{c}\text{\scriptsize{$i$}}\\ 
\text{\scriptsize{$\vee$}}\\ \frac12\end{array},0,\cdots,0),$$
where $\beta_i=\frac12$.
Then $\begin{pmatrix}\alpha \\ \beta \end{pmatrix}_{\Omega}$ 
is a non-singular odd half-period of $\mathrm{Jac}(\gamma)$.
\end{proposition}

\begin{proof}
It is elementarily shown by using the fact that singular half-periods have at 
least two nonzero entries in $\alpha$.
\end{proof}

Fix a non-singular odd half-period
$\begin{pmatrix}\alpha \\ \beta \end{pmatrix}_{\Omega}$,
and denote the corresponding (non-singular odd) 
theta characteristic by $\delta$.
Define the
half order differential $h_{\delta}(x)$ on $\gamma$ by 
$$h_\delta^2(x)=
\sum_{i=1}^g \frac{\partial \theta[\alpha,\beta]}
                  {\partial z_i}(0)\omega_i(x).$$
Here $h_{\delta}(x)$ is the holomorphic section 
of the line bundle corresponding to $\delta$. 
Then the prime form is defined by
\begin{align}\label{prime-form}
  E(x,y)=
  \frac{\theta[\alpha,\beta](\eta(y-x))}{h_{\delta}(x)h_{\delta}(y)}
\end{align}
for $x,y \in \gamma$.
We do not use $h_{\delta}(x)$ in this paper except in the following theorem.
See \cite{Fay73} or \cite{Mumford-Book} for general settings other than
the hyperelliptic case.

\begin{theorem}\cite[eq.(45)]{Fay73}
Let $\gamma$ be a hyperelliptic curve,
$\theta({\bf z})$ be the Riemann's theta function \eqref{R-theta}, 
and $E(x,y)$ be the prime form \eqref{prime-form}. 
Then for $P_1,P_2,P_3,P_4$ in the 
universal covering space of $\gamma$ and
${\bf z}\in \C^g$, the formula
\begin{align*}
&\theta({\bf z}+\int_{P_1}^{P_3})\theta({\bf z}+\int_{P_2}^{P_4})
E(P_3,P_2)E(P_1,P_4)\\
&+\theta({\bf z}+\int_{P_2}^{P_3})\theta({\bf z}+\int_{P_1}^{P_4})
E(P_3,P_1)E(P_4,P_2)\\
=\ &
\theta({\bf z}+\int_{P_1+P_2}^{P_3+P_4})\theta({\bf z})
E(P_3,P_4)E(P_1,P_2)
\end{align*}
holds, where $\int_{P_i}^{P_j}$ denotes $\eta(P_j-P_i)$.
\end{theorem}

By eliminating the common denominator,
we have 
\begin{align}\label{Fay-original}
\begin{split}
&\theta({\bf z}+\int_{P_1}^{P_3})\theta({\bf z}+\int_{P_2}^{P_4})
\theta[\alpha,\beta](\int_{P_3}^{P_2})\theta[\alpha,\beta](\int_{P_1}^{P_4})\\
&+\theta({\bf z}+\int_{P_2}^{P_3})\theta({\bf z}+\int_{P_1}^{P_4})
\theta[\alpha,\beta](\int_{P_3}^{P_1})\theta[\alpha,\beta](\int_{P_4}^{P_2})\\
=\ &
\theta({\bf z}+\int_{P_1+P_2}^{P_3+P_4})\theta({\bf z})
\theta[\alpha,\beta](\int_{P_3}^{P_4})\theta[\alpha,\beta](\int_{P_1}^{P_2}).
\end{split}
\end{align}

%%%%%%%%%%%%%%%%%%%%%%%%%%%%%%%
\subsection{Tropical analogue of Fay's identity}
%%%%%%%%%%%%%%%%%%%%%%%%%%%%%%%%%

For a positive definite symmetric matrix $K \in M_g(\R)$ and 
$\beta \in \R^g$ we define 
$$ 
  q_\beta({\bf m},{\bf Z}) = \frac12 {\bf m}K{\bf m}^{\bot}+
  {\bf m}({\bf Z}+\beta K)^{\bot}
  \qquad ({\bf Z} \in \R^g, ~ {\bf m} \in \Z^g),
$$
and write the tropical Riemann's theta function as
$$
  \Theta({\bf Z})=\min_{{\bf m}\in Z^g} q_0({\bf m},{\bf Z})
  \qquad ({\bf Z}\in \R^g).
$$ 
Let us introduce a generalization of the tropical Riemann's theta function:
$$\Theta[\beta]({\bf Z}) := \frac12 \beta K \beta^{\bot}+
 \beta {\bf Z}^{\bot}+
\min_{{\bf m} \in \Z^g} q_\beta({\bf m},{\bf Z}).$$ 
We write $\arg_{{\bf m} \in \Z^g} q_\beta({\bf m},{\bf Z})$ 
for ${\bf m}\in \Z^g$ where $q_\beta({\bf m},{\bf Z})$ takes the
minimum value.

\begin{proposition}\label{Theta-period}
The function $\Theta[\beta](\mathbf{Z})$ satisfies the following properties:
\\
{\rm (i)} the periodicity in $\beta$:
$$\Theta[\beta+{\bf l}]({\bf Z})=\Theta[\beta]({\bf Z})  
\quad ({\bf l}\in \Z^g),$$
{\rm (ii)} the quasi-periodicity in ${\bf Z}$:
$$\Theta[\beta]({\bf Z}+{\bf l}K)=-\frac12 {\bf l}K{\bf l}^{\bot}
-{\bf l}{\bf Z}^{\bot}
+\Theta[\beta]({\bf Z})  \quad ({\bf l}\in \Z^g),$$
{\rm (iii)} the symmetry in $\beta$ and $\mathbf{Z}$:
$$\Theta[\beta](-{\bf Z})=\Theta[-\beta]({\bf Z}).$$
{\rm (iv)} If $\beta \in \frac12 \Z^g/\Z^g$,  then $\Theta[\beta]({\bf Z})$ is 
an even function with respect to ${\bf Z}$ and $\beta$.
\end{proposition}
\begin{proof}
(i) and (ii) Replace ${\bf m}$ by ${\bf m}-{\bf l}$ in
$$\Theta[\beta+{\bf l}]({\bf Z})=
\frac12 (\beta+{\bf l})K(\beta+{\bf l})^{\bot}+(\beta+{\bf l}) {\bf Z}^{\bot} 
  +\min_{{\bf m} \in \Z^g}[\frac12 {\bf m}K{\bf m}^{\bot}
+{\bf m}({\bf Z}+(\beta+{\bf l})K)^{\bot}]$$
and
$$\Theta[\beta]({\bf Z}+{\bf l}K)=\frac12 \beta K\beta^{\bot}
+ \beta ({\bf Z}+{\bf l} K)^{\bot} 
  +\min_{{\bf m} \in \Z^g}[\frac12 {\bf m}K{\bf m}^{\bot}
+{\bf m}({\bf Z}+{\bf l} K+\beta K)^{\bot}].$$
(iii) Replace ${\bf m}$ by $-{\bf m}$ in
$$\Theta[\beta](-{\bf Z})=\frac12 \beta K\beta^{\bot}-\beta {\bf Z}^{\bot} 
  +\min_{{\bf m} \in \Z^g}[\frac12 {\bf m}K{\bf m}^{\bot}
+{\bf m}(-{\bf Z}+\beta K)^{\bot}].$$
(iv) By (i) and (ii), if $\beta \in \frac12 \Z^g/\Z^g$,  
we have $$\Theta[\beta](-{\bf Z})=\Theta[-\beta]({\bf Z})
=\Theta[-\beta+2\beta]({\bf Z})=\Theta[\beta]({\bf Z}).$$
\end{proof}

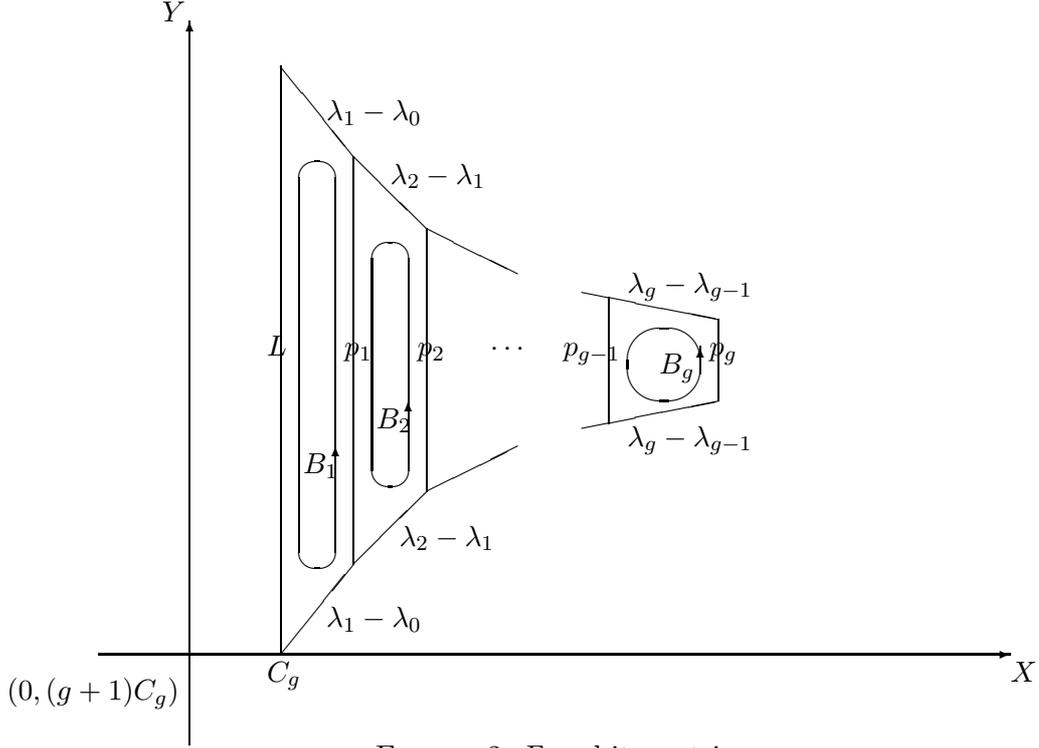
\begin{figure}
\begin{center}
\unitlength=1.2mm
\begin{picture}(100,70)(0,5)
\put(0,10){\line(1,0){20}}
\put(20,10){\vector(1,0){80}}
\put(100,7){$X$}
\put(10,0){\line(0,1){10}}
\put(10,10){\line(0,1){65}}
\put(10,75){\vector(0,1){5}}
\put(7,80){$Y$}
\put(-10,5){$(0,(g+1)C_g)$}

\put(25,13){$\lambda_1-\lambda_0$}
\put(33,22){$\lambda_2-\lambda_1$}
\put(58,33){$\lambda_g-\lambda_{g-1}$}

\put(25,69){$\lambda_1-\lambda_0$}
\put(32,62){$\lambda_2-\lambda_1$}
\put(58,50){$\lambda_g-\lambda_{g-1}$}

\put(20,10){\line(4,5){8}}
\put(28,20){\line(1,1){8}}
\put(36,28){\line(2,1){10}}
\put(53,35){\line(5,1){15}}

\put(20,75){\line(4,-5){8}}
\put(28,65){\line(1,-1){8}}
\put(36,57){\line(2,-1){10}}
\put(53,50){\line(5,-1){15}}

\put(20,10){\line(0,1){65}}
\put(28,20){\line(0,1){45}}
\put(36,28){\line(0,1){29}}
\put(56,35.5){\line(0,1){14}}
\put(68,38){\line(0,1){9}}

\put(18.5,43){$L$}
\put(18.5,7){$C_g$}
\put(27,43){$p_1$}
\put(35,43){$p_2$}
\put(43,43){$\cdots$}
\put(51,43){$p_{g-1}$}
\put(67,43){$p_{g}$}

\put(24,42){\oval(4,45)}
\put(26,30){\vector(0,1){3}}
\put(22.5,30){$B_1$}
\put(32,42){\oval(4,27)}
\put(34,35){\vector(0,1){3}}
\put(30.5,35){$B_2$}
\put(62,42){\oval(8,8)}
\put(66,41){\vector(0,1){3}}
\put(61.5,41){$B_g$}

\end{picture}
\caption{$\Gamma$ and its metric}\label{Gamma-metric}
\end{center}
\end{figure}

In the rest of this section,
let us restrict ourselves to the case of a family of 
hyperelliptic tropical curves given by \eqref{trop-curve}
with the generic condition \eqref{CD-condition}.
We define $L$, $\lambda_0,\lambda_1,\cdots,\lambda_g$ and 
$p_1,\cdots,p_g$ by 
\begin{align}
  \label{partition}
  \begin{split}
  &L = C_{-1}-2(g+1)C_g, \qquad \lambda_0=C_g, 
  \\
  &\lambda_i = C_{g-i}-C_{g-i+1},
  \qquad
  p_i = L - 2 \sum_{j=1}^g \min[\lambda_i-\lambda_0,\lambda_j-\lambda_0]
  \quad (\mbox{for}\ 1 \leq i \leq g,).
  \end{split}
\end{align}
We set $\vec{\lambda}=(\lambda_1-\lambda_0, \dots, \lambda_g-\lambda_{g-1})$.
Due to the condition \eqref{CD-condition} one sees 
\begin{align}
  \begin{split}
  &\lambda_0 < \lambda_1 < \lambda_2 < \cdots < \lambda_g,
  \qquad
  0< p_g < p_{g-1} < \cdots < p_1, 
  \qquad
  2 \sum_{i=1}^g (\lambda_i -\lambda_0) < L, 
  \\
  &\vec{g} \cdot \vec{\lambda}
  =\sum_{i=1}^g (\lambda_i -\lambda_0)=C_0-(g+1)C_g \label{g-lambda}.
  \end{split}
\end{align}
We show the maximal compact subset $\Gamma$ of $\tilde{\Gamma}$
in Figure \ref{Gamma-metric},
where a scalar on each edge denotes its length$_d$.
The period matrix $K = (K_{ij})$ \eqref{periodK} for $\Gamma$ becomes
\begin{align}\label{hyperell-periodK}
    K_{ij} = 
    \begin{cases}
       p_{i-1}+p_i+2(\lambda_i-\lambda_{i-1}) >0, \text{ where } p_0=L 
       \quad (i=j) \\
        -p_i < 0 \quad (j=i+1) \\
       -p_j <0 \quad (i=j+1)\\
       0 \quad (\text{otherwise}).
    \end{cases}
\end{align}

\begin{theorem}{\rm (Tropical analogue of Fay's trisecant identity)}
\label{tropicalFay}
Let $\tilde{\Gamma}$ be a smooth tropical curve 
given by \eqref{trop-curve} with the generic condition
\eqref{CD-condition}.

For $\beta \in \frac12 \Z^g$ $(\beta \neq 0 \mod \Z^g)$, 
set $\alpha\in\frac12 \Z^g$ as
$$\alpha=(0,\cdots,0,\textstyle -\frac12, \begin{array}[b]{c}\text{\scriptsize{$i$}}\\ 
\text{\scriptsize{$\vee$}}\\ \frac12\end{array}, \textstyle 
0,\cdots,0),$$
where $\beta_j=0$ for $1\leq j \leq i-1$ and $\beta_i\neq 0$.
For $P_1, P_2, P_3, P_4$ 
on the universal covering space of $\Gamma$, 
we define the sign $s_i\in \{\pm 1\}$ $(i=1,2,3)$ as 
$s_i=(-1)^{k_i}$,
where
\begin{align*}
k_1&=2\alpha\cdot 
\left(\arg_{{\bf m} \in \Z^g} q_\beta({\bf m},\int_{P_3}^{P_2})+
\arg_{{\bf m} \in \Z^g} q_\beta({\bf m},\int_{P_1}^{P_4})\right),\\
k_2&=2\alpha\cdot 
\left(\arg_{{\bf m} \in \Z^g} q_\beta({\bf m},\int_{P_3}^{P_1})+
\arg_{{\bf m} \in \Z^g} q_\beta({\bf m},\int_{P_4}^{P_2})\right),\\
k_3&=1+2\alpha\cdot 
\left(\arg_{{\bf m} \in \Z^g} q_\beta({\bf m},\int_{P_3}^{P_4})+
\arg_{{\bf m} \in \Z^g} q_\beta({\bf m},\int_{P_1}^{P_2})\right).
\end{align*}
For ${\bf Z} \in \R^g$, set  $F_1, F_2, F_3 \in \R$ as
\begin{align}\label{trop-fay}
\begin{split}
F_1&=\Theta({\bf Z}+\int_{P_1}^{P_3})+\Theta({\bf Z}+\int_{P_2}^{P_4})
 +\Theta[\beta](\int_{P_3}^{P_2})+\Theta[\beta](\int_{P_1}^{P_4}),\\
F_2&=\Theta({\bf Z}+\int_{P_2}^{P_3})+\Theta({\bf Z}+\int_{P_1}^{P_4})
 +\Theta[\beta](\int_{P_3}^{P_1})+\Theta[\beta](\int_{P_4}^{P_2}),\\
F_3&=\Theta({\bf Z}+\int_{P_1+P_2}^{P_3+P_4})+\Theta({\bf Z})
+\Theta[\beta](\int_{P_4}^{P_3})+\Theta[\beta](\int_{P_1}^{P_2}).
\end{split}
\end{align}
Then, the formula
$$F_i=\min[F_{i+1},F_{i+2}]$$  
holds if $s_i=\pm 1,s_{i+1}=s_{i+2}=\mp 1$ for $i \in \Z / 3 \Z$.
\end{theorem}

\begin{remark}
(i) The case of $s_1=s_2=s_3$ does not occur. \\
(ii) The sign $s_i$ is not determined
when the corresponding 
$\arg_{{\bf m} \in \Z^g} q_\beta({\bf m},\int_{P_j}^{P_k})$
is not unique.
Sometimes it is possible to    
move the points $P_j$'s slightly so that $s_i$ is determined.
But it can not be done always. (See Example~\ref{ex-counter}.)
\end{remark}

The following theorem is the key to the proof of 
Theorem \ref{tropicalFay},
which links the Abel-Jacobi map on $\gamma$ to that on $\Gamma$:

\begin{theorem}\label{UDAJ}
By the UD-limit with the scale transformation
\begin{align}\label{xyzomega-trans}
  |x|=e^{-X/\ve}, \ |y|=e^{-Y/\ve},\ 
  {\bf z}=-\frac{{\bf Z}}{2\pi \I \ve}, \quad 
  \Omega_{ij}=-\frac{\tilde{K}_{ij}}{2\pi \I \ve},
\end{align}  
the Abel-Jacobi map \eqref{A-J} on $\gamma$
%$\sum_i {P_i}\mapsto \sum_i(\int_{P_0}^{P_i}\omega_j)_j$
becomes the tropical Abel-Jacobi map \eqref{UD-AJ} on $\Gamma$ as
$$\eta: \Div(\Gamma)\to J(\Gamma); ~\sum_i m_i{P_i}\mapsto 
\sum_i m_i (\<\Gamma_{P_0}^{P_i},\tilde{B}_j \>)_{j=1,\cdots,g},$$
where $P_0 \in \Gamma$ is a base point and  
$\tilde{B}_j=B_{g-j+1}+B_{g-j+2}+\cdots+B_g$.
In this limit, $\tilde{K}$ becomes
$\tilde{K}_{ij}=\<\tilde{B}_i,\tilde{B}_j\>$.
\end{theorem}
 
We will prove it analytically as a variation of the result in 
\cite{IwaoTokihiro07}. Since it is straightforward but tedious, we give
it in the appendix. 

\begin{remark}
(i) The cycles $\tilde{B}_j$'s are obtained from $B_j$'s
by the base change as $(\Tilde{B}_1,\cdots,\Tilde{B}_g) = (B_1,\cdots,B_g) T$
with 
$$
  T=\begin{pmatrix} 0&&1\\ &\iddots& \vdots \\[2mm] 1&\dots &1 
  \end{pmatrix} \in M_g(\C).
$$
Thus the tropical Abel-Jacobi map \eqref{UD-AJ} 
is obtained from the complex Abel-Jacobi map \eqref{A-J}
through the UD-limit.\\
(ii) We have $\tilde{K}=TKT$
which corresponds to the base change of $H_1(\gamma,\Z)$ as 
$(a_i)_i \mapsto (a_i)_i \, T $ and 
$(b_i)_i \mapsto (b_i)_i \,T^{-1}= (b_{g-i+1}-b_{g-i})_i$.  
Thus 
$\alpha I + \beta \Omega$ of Theorem~\ref{tropicalFay} corresponds to 
$\alpha T + \beta T^{-1} \Omega$
which is a nonsingular odd half period of $\mathrm{Jac}(\gamma)$ 
from Proposition~\ref{nonsing.odd}.
\end{remark}

\begin{proof}[Proof of Theorem \ref{tropicalFay}]
By changing the basis of $H_1(\gamma, \Z)$ from
$(a_i)_i, (b_i)_i$ to $(a_i)_i \, T, (b_i)_i \, T^{-1}$, 
the limit of $(b_i)_i \, T^{-1}$ becomes $(B_i)_i$.
By the scale transformation of ${\bf z}$ and $\Omega$ as \eqref{xyzomega-trans}
the theta function $\theta[\alpha,\beta]({\bf z})$ becomes
\begin{align} \label{theta-scale}
\begin{split}
\theta[\alpha,\beta]({\bf z}) =
e^{2\pi \I \beta \alpha^{\bot}} & \exp \{-\frac{1}{\ve} 
(\frac12\beta K \beta^{\bot}+ \beta {\bf Z}^{\bot} )\} \\
 & \times
\sum_{{\bf m}\in \Z^g} e^{2 \pi \I {\bf m} \alpha^{\bot}} 
\exp \{-\frac{1}{\ve}(\frac12 {\bf m}K {\bf m}^{\bot} 
+{\bf m}({\bf Z}+K \beta )^{\bot})\}.
\end{split}
\end{align}

Step 1:
Since $e^{4\pi \I \beta \alpha^{\bot}}=\pm 1$ is a common factor and 
$e^{2\pi \I {\bf m} \alpha^{\bot}}=\pm 1$, we can set 
\begin{align}\label{f1}
f_1:=e^{-4\pi \I \beta \alpha^{\bot}} 
\theta({\bf z}+\int_{P_1}^{P_3})\theta({\bf z}+\int_{P_2}^{P_4})
\theta[\alpha,\beta](\int_{P_3}^{P_2})\theta[\alpha,\beta](\int_{P_1}^{P_4})
=f_1^{+}-f_1^{-},
\end{align}
where we take $f_1^{+}>0$ and $f_1^{-}>0$ as the part of 
$e^{2\pi \I {\bf m} \alpha^{\bot}}= 1$ 
and of $e^{2\pi \I {\bf m} \alpha^{\bot}}= -1$ respectively. Similarly we set
\begin{align*}
f_2:=&e^{-4\pi \I \beta \alpha^{\bot}}\theta({\bf z}+\int_{P_2}^{P_3})
\theta({\bf z}+\int_{P_1}^{P_4})
\theta[\alpha,\beta](\int_{P_3}^{P_1})\theta[\alpha,\beta](\int_{P_4}^{P_2})
=f_2^{+}-f_2^{-},\\
f_3:=&e^{-4\pi \I \beta \alpha^{\bot}}
\theta({\bf z}+\int_{P_1+P_2}^{P_3+P_4})\theta({\bf z})
\theta[\alpha,\beta](\int_{P_3}^{P_4})\theta[\alpha,\beta](\int_{P_1}^{P_2})
=f_3^{+}-f_3^{-}.
\end{align*}
Then $f_1^{+}+f_2^{+}+f_3^{-}=f_1^{-}+f_2^{-}+f_3^{+}$ holds 
from \eqref{Fay-original}.
We write $F_i^{\pm}$ for the UD-limit of $f_i^{\pm}$.
Then we obtain $F_i=\min[F_i^+,F_i^-]$ and 
$\min[F_1^+,F_2^+,F_3^-]=\min[F_1^-,F_2^-,F_3^+]$.\\
Now we have the following cases:\\
(i) If $F_1^+<F_1^-$, $F_2^+<F_2^-$ and $F_3^+<F_3^-$, then 
$\min[F_1,F_2]=F_3$.\\
(ii) If $F_1^+>F_1^-$, $F_2^+>F_2^-$ and $F_3^+>F_3^-$, then 
$\min[F_1,F_2]=F_3$.\\
(iii) If $F_1^+<F_1^-$, $F_2^+>F_2^-$ and $F_3^+<F_3^-$, then 
$\min[F_2,F_3]=F_1$.\\
(iv) If $F_1^+>F_1^-$, $F_2^+<F_2^-$ and $F_3^+>F_3^-$, then 
$\min[F_2,F_3]=F_1$.\\
(v) If $F_1^+<F_1^-$, $F_2^+>F_2^-$ and $F_3^+>F_3^-$, then 
$\min[F_3,F_1]=F_2$.\\
(vi) If $F_1^+>F_1^-$, $F_2^+<F_2^-$ and $F_3^+<F_3^-$, then 
$\min[F_3,F_1]=F_2$.\\
(vii) If $F_1^+<F_1^-$, $F_2^+<F_2^-$ and $F_3^+>F_3^-$, then 
$f_1^++f_2^++f_3^- > f_1^-+f_2^-+f_3^+$ for sufficiently small $\ve$, which is
a contradiction. \\
(viii) If $F_1^+>F_1^-$, $F_2^+>F_2^-$ and $F_3^+<F_3^-$, then 
$f_1^++f_2^++f_3^- < f_1^-+f_2^-+f_3^+$ for sufficiently small $\ve$, which is
a contradiction. 

Step 2:
We check the definition of $s_i$. By \eqref{theta-scale} and \eqref{f1},
$s_1=1$ means
$$e^{\pi \I \alpha \cdot \arg_{{\bf m} \in \Z^g} 
q_0({\bf m},\int_{P_3}^{P_2})}\cdot
e^{\pi \I \alpha \cdot \arg_{{\bf m} \in \Z^g} 
q_0({\bf m},\int_{P_1}^{P_4})} =1,$$
and thus $F_1=F_1^+$.
Similarly we have $F_1=F_1^-$ if $s_1=-1$, 
$F_2=F_2^+$ if $s_2=1$, $F_2=F_2^-$ if $s_2=-1$,
$F_3=F_3^+$ if $s_3=-1$ and $F_3=F_3^-$ if $s_3=1$.

From Steps 1 and 2 the claim follows.
%This fact together with (i),$\dots$,(viii) above we have the claim.
\end{proof}

%%%%%%%%%%%%%%%%%%%%%%%%%%%%%%%%%%%%%%%%%%%
\subsection{Bilinear equation of Toda type}
%%%%%%%%%%%%%%%%%%%%%%%%%%%%%%%%%%%%%%%%%%%

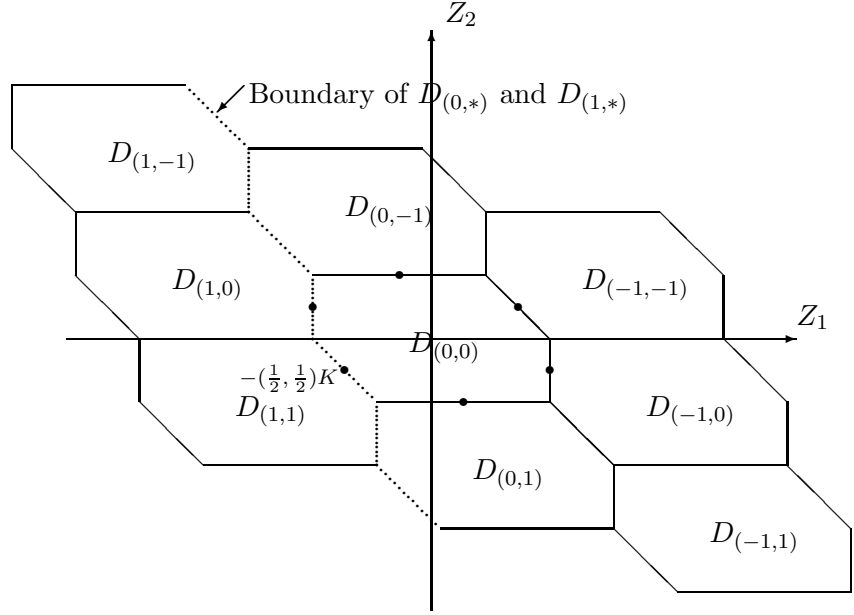
\begin{figure}
\begin{center}
\unitlength=0.6mm
\begin{picture}(160,120)(-80,-60)
\put(0,0){\vector(1,0){80}}
\put(0,0){\line(-1,0){80}}
\put(0,0){\vector(0,1){68}}
\put(0,0){\line(0,-1){60}}
\put(80,3){$Z_1$}
\put(3,70){$Z_2$}

%D_{(0,0)}
\put(26,0){\line(0,-1){14}}
\put(26,0){\line(-1,1){14}}
\put(-26,14){\line(1,0){38}}
\multiput(-26, 0)(0, 1){14}{\circle*{0.2}}
%\put(-26,0){\line(0,1){14}}
\multiput(-26, 0)(1, -1){14}{\circle*{0.2}}
%\put(-26,0){\line(1,-1){14}}
\put(26,-14){\line(-1,0){38}}
\put(-5,-3){$D_{(0,0)}$}
\put(-26,7){\circle*{2}}
\put(26,-7){\circle*{2}}
\put(-7,14){\circle*{2}}
\put(7,-14){\circle*{2}}
\put(19,7){\circle*{2}}
\put(-19,-7){\circle*{2}}
\put(-42,-10){{\scriptsize $-(\frac12,\frac12)K$}}

%D_{(1,0)}
%\put(-26,14){\line(0,-1){14}}
%\put(-26,14){\line(-1,1){14}}
\put(-78,28){\line(1,0){38}}
\put(-78,14){\line(0,1){14}}
\put(-78,14){\line(1,-1){14}}
\put(-26,0){\line(-1,0){38}}
\put(-57,11){$D_{(1,0)}$}

%D_{(1,-1)}
%\put(-40,42){\line(0,-1){14}}
\multiput(-40, 42)(-1, 1){14}{\circle*{0.2}}
%\put(-40,42){\line(-1,1){14}}
\put(-92,56){\line(1,0){38}}
\put(-92,42){\line(0,1){14}}
\put(-92,42){\line(1,-1){14}}
\put(-40,28){\line(-1,0){38}}
\put(-71,39){$D_{(1,-1)}$}
\put(-40,52){Boundary of $D_{(0,*)}$ and $D_{(1,*)}$} 
\put(-41,56){\vector(-1,-1){6}}

%D_{(1,1)}
%\put(-12,-14){\line(0,-1){14}}
%\put(-12,-14){\line(-1,1){14}}
\put(-64,0){\line(1,0){38}}
\put(-64,-14){\line(0,1){14}}
\put(-64,-14){\line(1,-1){14}}
\put(-12,-28){\line(-1,0){38}}
\put(-43,-17){$D_{(1,1)}$}

%D_{(0,1)}
\put(40,-28){\line(0,-1){14}}
\put(40,-28){\line(-1,1){14}}
\put(-12,-14){\line(1,0){38}}
\multiput(-12, -28)(0, 1){14}{\circle*{0.2}}
%\put(-12,-28){\line(0,1){14}}
\multiput(-12, -28)(1, -1){14}{\circle*{0.2}}
%\put(-12,-28){\line(1,-1){14}}
\put(40,-42){\line(-1,0){38}}
\put(9,-31){$D_{(0,1)}$}

%D_{(-1,0)}
\put(78,-14){\line(0,-1){14}}
\put(78,-14){\line(-1,1){14}}
\put(26,0){\line(1,0){38}}
\put(26,-14){\line(0,1){14}}
\put(26,-14){\line(1,-1){14}}
\put(78,-28){\line(-1,0){38}}
\put(47,-17){$D_{(-1,0)}$}

%D_{(-1,1)}
\put(92,-42){\line(0,-1){14}}
\put(92,-42){\line(-1,1){14}}
\put(40,-28){\line(1,0){38}}
\put(40,-42){\line(0,1){14}}
\put(40,-42){\line(1,-1){14}}
\put(92,-56){\line(-1,0){38}}
\put(61,-45){$D_{(-1,1)}$}

%D_{(-1,-1)}
\put(64,14){\line(0,-1){14}}
\put(64,14){\line(-1,1){14}}
\put(12,28){\line(1,0){38}}
\put(12,14){\line(0,1){14}}
\put(12,14){\line(1,-1){14}}
\put(64,0){\line(-1,0){38}}
\put(33,11){$D_{(-1,-1)}$}

%D_{(0,-1)}
\put(12,28){\line(0,-1){14}}
\put(12,28){\line(-1,1){14}}
%\put(-20,22){\circle*{2}}
%\put(-27,20){{\scriptsize ${\bf Z}_0$}}
%\multiput(-20, 22)(2, 0){16}{\line(1,0){1}}
%\put(-18, 17){{\tiny dist. to Wall}}
\put(-40,42){\line(1,0){38}}
\multiput(-40, 28)(0, 1){14}{\circle*{0.2}}
%\put(-40,28){\line(0,1){14}}
\multiput(-40, 28)(1, -1){14}{\circle*{0.2}}
%\put(-40,28){\line(1,-1){14}}
\put(12,14){\line(-1,0){38}}
\put(-19,27){$D_{(0,-1)}$}

\end{picture}
\caption{Fundamental regions of $g=2$ ~(${\bf Z} = (Z_1,Z_2)$)}\label{Fig_D_n}
\end{center}
\end{figure}

We return to the definition of the tropical Riemann's theta function and
investigate the fundamental regions.
We define 
the fundamental region $D_{{\bf m}}$ of $\Theta({\bf Z})$ as
$D_{{\bf m}}=\{{\bf Z}\in \R^g ~|~ \Theta({\bf Z})=
\frac12 {\bf m}K{\bf m}^{\bot}+m{\bf Z}^{\bot} \}$, 
which is explicitly written as 
\begin{align*}
  D_{{\bf m}}=\{{\bf Z}\in \R^g ~|~ -{\bf l} {\bf Z}^{\bot} \leq  
{\bf l}K({\bf m}+\frac12 {\bf l})^{\bot}\
  \mbox{ for any ${\bf l} \in \Z^g$} \}. 
\end{align*} 
Note that $\arg_{{\bf m}\in \Z^g} q_0({\bf m},{\bf Z})={\bf m}$ if and only if
${\bf Z}$ is in the interior of $D_{{\bf m}}$. 
See Figure \ref{Fig_D_n} for the $g=2$ case.

We easily see the following:
\begin{lemma}\label{K-etc}
The period matrix $K$ \eqref{hyperell-periodK}
satisfies the following properties:
\begin{align*}
  &\mathrm{(i)}~ 
  \mathbb{I} K
   = p_g {\bf e}_g+ L {\bf e}_1+ 2 \vec{\lambda},
  \qquad 
  \mathrm{(ii)} ~\sum_{j=1}^g K_{ij} > 0, 
  \qquad 
  \mathrm{(iii)} ~\vec{g}K=(g+1)L{\bf e}_1.
\end{align*}
\end{lemma}

\begin{lemma}\label{lemma:Dm-domain}
$D_{{\bf m}}$ is written as
\begin{align}\label{region}
  D_{{\bf m}}=&\{{\bf Z}\in \R^g ~|~ -{\bf l} {\bf Z}^{\bot} \leq  
{\bf l}K({\bf m}+\frac12 {\bf l})^{\bot}\\
&\quad  \mbox{ for any ${\bf l}=\pm({\bf e}_j+{\bf e}_{j+1}+\cdots+{\bf e}_k)$ 
 such that $1\leq j\leq k \leq g$} \}. \nonumber
\end{align} 
\end{lemma}

\begin{proof}
Since $D_{{\bf m}}=D_{{\bf 0}}-{\bf m}K$ from the definition, 
it is enough to show for $D_{{\bf 0}}$.
We show that if ${\bf Z} \in \R^g$ satisfies
\begin{align*}
-{\bf l} {\bf Z}^{\bot} > &  \frac{1}{2}{\bf l}K{\bf l}^{\bot}
\end{align*}
for some ${\bf l} \in \Z^g$, then there exists 
${\bf l}'=\pm({\bf e}_j+\cdots+{\bf e}_k)$ 
for $1\leq j\leq k \leq g$, which satisfies 
\begin{align}\label{region2}
-{\bf l}' {\bf Z}^{\bot} > &  \frac{1}{2}{\bf l}'K({\bf l}')^{\bot}.
\end{align}
For a vector ${\bf v} \in \R^g$, 
let ${\bf v} \geq 0$ denote that all elements of ${\bf v}$ are 
greater than or equal to zero. 

This lemma is shown by checking the following three claims (a)-(c):
\\
(a) ~{\it There exists ${\bf l}'\geq 0$ or ${\bf l}'\leq 0$ which
satisfies \eqref{region2}.}
Decompose $l$ as ${\bf l}={\bf l}_++{\bf l}_-$ such that
${\bf l}_+ \geq 0 $, ${\bf l}_- \leq 0$ and ${\bf l}_+\cdot {\bf l}_-=0$.
Then we have
$$-({\bf l}_++{\bf l}_-){\bf Z}^{\bot}> 
\frac12 ({\bf l}_++{\bf l}_-)K({\bf l}_++{\bf l}_-)^{\bot}
= \frac12 ({\bf l}_+K{\bf l}_+^{\bot} 
+ {\bf l}_-K{\bf l}_-^{\bot})+{\bf l}_+K{\bf l}_-^{\bot} .$$
Since $K_{ij}\leq 0$ for $i \neq j$, 
$$ \frac12 {\bf l}_+ K {\bf l}_-^{\bot} \geq 0 $$
holds. We have
$$-({\bf l}_++{\bf l}_-){\bf Z}^{\bot} > 
\frac12 ({\bf l}_+K{\bf l}_+^{\bot} + {\bf l}_-K{\bf l}_-^{\bot})$$
and hence
$-{\bf l}_+{\bf Z}^{\bot}> 
\frac12 {\bf l}_+K{\bf l}_+^{\bot}$ or $-{\bf l}_-{\bf Z}^{\bot}
> \frac12 {\bf l}_-K{\bf l}_-^{\bot}$. 
\\
(b) ~{\it If ${\bf l}\geq 0$ or ${\bf l}\leq 0$, then 
there exists ${\bf l}'\in \pm\{0,1\}^g$ which
satisfies \eqref{region2}.}
Without loss of generality we can assume that ${\bf l}\geq 0$.
For simplicity we also assume that ${\bf l}=(l_1,l_2,\dots,l_g)$ satisfies 
$l_1\geq l_2 \geq \cdots \geq l_g$. One can prove similarly in other cases. 
We use the induction on $k:=l_1$. Assume that 
if $l_1^*\leq k-1$ for ${\bf l}^*$ satisfying the above assumptions 
for {\bf l}, then 
there exists ${\bf l}'\in \{0,1\}^g$
satisfying \eqref{region2}.
We set a natural number $r$ as 
$l_1=l_2=\cdots =l_r>l_{r+1}\geq \cdots \geq l_g$ and a vector
${\bf l}^*$ as 
${\bf l}^*={\bf l} -(l_1-l_{r+1}){\mathbb I}_r$.
Then ${\bf l}$ is decomposed as 
${\bf l}=(l_1-l_{r+1}){\mathbb I}_r+{\bf l}^*$ and we have
\begin{align*}
-{\bf l}{\bf Z}^{\bot}>& \frac12 {\bf l}K{\bf l}^{\bot}\\ 
=& 
\frac12 (l_1-l_{r+1})^2{\mathbb I}_r K {\mathbb I}_r^{\bot} 
+ \frac12 {\bf l}^*K ({\bf l}^*)^{\bot}
+ (l_1-l_{r+1}) {\mathbb I}_r K ({\bf l}^*)^{\bot},
\end{align*}
where  
\begin{align*}
{\mathbb I}_r K({\bf l}^*)^{\bot}=&
{\mathbb I}_r \cdot \left(
\begin{array}{c}
l_{r+1}K_{11}+\cdots+l_{r+1}K_{1r}+l_{r+1}K_{1 \,r+1}+\cdots+l_gK_{1 g}\\
\vdots\\
l_{r+1}K_{r1}+\cdots+l_{r+1}K_{rr}+l_{r+1}K_{r\,r+1}+\cdots+l_gK_{r g}
\\[2mm]
*\\
\vdots\\[1mm]
*
\end{array} 
\right)\\
\geq & l_{r+1} ((K_{11}+\cdots+K_{1g})+\cdots+(K_{r1}+\cdots+K_{rg}))\\
\geq& 0.
\end{align*}
Thus we obtain
$$ -((l_1-l_{r+1}){\mathbb I}_r+{\bf l}^*){\bf Z}^{\bot} 
> \frac12 (l_1-l_{r+1})^2 {\mathbb I}_r K {\mathbb I}_r^{\bot}
+\frac12 {\bf l}^*K ({\bf l}^*)^{\bot},
$$ 
and therefore
$$- {\mathbb I}_r {\bf Z}^{\bot}> \frac12 {\mathbb I}_r K {\mathbb I}_r^{\bot}
\quad \mbox{or} \quad
- {\bf l}^* {\bf Z}^{\bot}>  \frac12 {\bf l}^*K ({\bf l}^*)^{\bot}
$$
since $l_1-l_{r+1}\geq 1$. 
If the former holds, we obtain the claim by 
setting ${\bf l}'={\mathbb I}_r$. 
Otherwise, since $l_1^* \leq k-1$, 
there exists ${\bf l}'\in \{0,1\}^g$ satisfying \eqref{region2}
by the induction hypothesis.
\\
(c) ~ {\it Suppose ${\bf l} \geq 0$ or ${\bf l} \leq 0$ and $l_s=0$ 
for some $1<s<g$,
and decompose ${\bf l}$ as ${\bf l}={\bf l}_L+{\bf l}_R$ 
in such a way that
$({\bf l}_L)_i=0$ for $s\leq i\leq g$
and $({\bf l}_R)_i=0$ for $1 \leq i\leq s$.
Then ${\bf l}'={\bf l}_L$ or ${\bf l}'={\bf l}_R$
satisfies \eqref{region2}.}
We can assume ${\bf l}\geq 0$. We have 
\begin{align*}
-({\bf l}_L+{\bf l}_R){\bf Z}^{\bot}> & 
\frac12 ({\bf l}_L+{\bf l}_R)K({\bf l}_L+L_R)^{\bot}\\ 
=&\frac12 {\bf l}_L K {\bf l}_L^{\bot}
+ \frac12 {\bf l}_R K {\bf l}_R^{\bot} +{\bf l}_L K {\bf l}_R^{\bot},
\end{align*}
where 
$$
{\bf l}_L K {\bf l}_R^{\bot}= \sum_{i=1}^{s-1} \sum_{j=s+1}^g l_il_j K_{ij} =0.
$$
Thus the claim follows.
\end{proof}

\begin{proposition}\label{bilinear-theta}
The tropical Riemann's theta function satisfies
\begin{align*}
&\min[2\Theta({\bf Z}+\vec{\lambda}), 
\Theta({\bf Z}-L{\bf e}_1)+\Theta({\bf Z}+L{\bf e}_1+2\vec{\lambda})+L]
=\Theta({\bf Z}+2\vec{\lambda})+\Theta({\bf Z}).
\end{align*}
\end{proposition}

\begin{proof}
Take $\beta=-\frac12{\mathbb I}$ and 
\begin{align*}
&P_1=(\lambda_g, \frac12 L+ \frac12 p_g+(g+1)C_g), \quad
P_2=(\lambda_0,(g+1)C_g),\\ 
&P_3=(\lambda_0,L+(g+1)C_g), \quad 
P_4=(\lambda_g, \frac12 L- \frac12 p_g+(g+1)C_g).
\end{align*}
Then we have 
\begin{align*}
&\int_{P_1}^{P_3}=\int_{P_2}^{P_4}=\vec{\lambda}, \quad
\int_{P_2}^{P_3}=-L{\bf e}_1, \\
&\int_{P_1}^{P_4}=L{\bf e}_1+2\vec{\lambda}, \quad
\int_{P_3}^{P_4}=\int_{P_1}^{P_2}=L{\bf e}_1+\vec{\lambda}. 
\end{align*}
We use the following lemma which will be proved after this proof.

\begin{lemma}\label{pointsin} \ Set $\beta = -\frac12{\mathbb I}$.
Then the followings are satisfied:
\\
{\rm (i)} 
the point ${\bf Z}=\beta K$ is on the boundary $\partial D_{{\bf 0}}$; \\
{\rm (ii)} all of
${\bf Z}=\beta K + nL{\bf e}_1+t \vec{\lambda}~ (n=0,1,t=0,1,2)$ except
$\beta K$ are in the interior of
$D_{{\bf 0}}$;
\\
{\rm (iii)} if $\beta K+ {\bf v}$ is in the interior of $D_{{\bf 0}}$, 
then $\beta K-{\bf v}$ is in the interior of $D_{\mathbb I}$.
\end{lemma} 

From (ii) and (iii), we have
$\arg_{{\bf m} \in \Z^g}q_\beta({\bf m}, nL{\bf e}_1+t \vec{\lambda})= {\bf 0}$
and
$\arg_{{\bf m} \in \Z^g}q_\beta({\bf m},-nL{\bf e}_1-t \vec{\lambda})
={\mathbb I}$
for $n=0,1,t=0,1,2$ except for $n=t=0$.
Further, from the definition of $\Theta[\beta]({\bf Z})$, 
we have the following:
\begin{eqnarray*}
&\Theta[\beta](\int_{P_3}^{P_2})=-\frac12 L+\frac12\beta K\beta^{\bot}, \quad&
\arg_{{\bf m} \in \Z^g}q_\beta({\bf m},\int_{P_3}^{P_2})= {\bf 0}, \\
&\Theta[\beta](\int_{P_1}^{P_4})=-\frac12L-(\lambda_g-\lambda_0)
+\frac12\beta K\beta^{\bot},\quad& 
\arg_{{\bf m} \in \Z^g}q_\beta({\bf m},\int_{P_1}^{P_4})={\bf 0},  \\
&\Theta[\beta](\int_{P_3}^{P_1})=\Theta[\beta](\int_{P_1}^{P_3})
=-\frac12 (\lambda_g-\lambda_0)+\frac12\beta K\beta^{\bot}, \quad&
\arg_{{\bf m} \in \Z^g}q_\beta({\bf m},\int_{P_3}^{P_1})={\mathbb I}, \\
&\Theta[\beta](\int_{P_4}^{P_2})=\Theta[\beta](\int_{P_2}^{P_4})
=-\frac12 (\lambda_g-\lambda_0)+\frac12\beta K\beta^{\bot}, \quad&
\arg_{{\bf m} \in \Z^g}q_\beta({\bf m},\int_{P_4}^{P_2})={\mathbb I}, \\ 
&\Theta[\beta](\int_{P_3}^{P_4})
=-\frac12 L- \frac12 (\lambda_g-\lambda_0)+\frac12\beta K\beta^{\bot},
\quad& \arg_{{\bf m} \in \Z^g}q_\beta({\bf m},\int_{P_3}^{P_4})={\bf 0},   \\
&\Theta[\beta](\int_{P_1}^{P_2})
=-\frac12 L- \frac12 (\lambda_g-\lambda_0) +\frac12\beta K\beta^{\bot},
\quad& \arg_{{\bf m} \in \Z^g}q_\beta({\bf m},\int_{P_1}^{P_2})={\bf 0},  
\end{eqnarray*}
thus $s_1=1$, $s_2=1$ and  $s_3=-1$ hold.
Substituting them into the tropical Fay's identity 
\eqref{trop-fay}, we obtain the claim.
\end{proof}

\begin{proof}[Proof of Lemma~\ref{pointsin}] \ 
In this proof we use
$$
  f_{\bf m}({\bf Z}, {\bf l}) := 
  {\bf l} {\bf Z}^{\bot} + {\bf l}K({\bf m}+\frac12 {\bf l})^{\bot},
$$
which is ``r.h.s - l.h.s'' of 
the conditional equation for $D_{\bf m}$ \eqref{region}.
\\
(i) For ${\bf l}={\mathbb I}$ and ${\bf m} = {\bf 0}$, we have 
$f_{\bf 0}(\beta K, \mathbb{I}) = (-{\mathbb I})(\frac12 K {\mathbb I})^{\bot}
+{\mathbb I}(\frac12 K {\mathbb I})^{\bot}=0$. 
For ${\bf l}\in \{0,1\}^g\setminus \{{\bf 0}, {\mathbb I}\}$, 
we have 
$$f_{\bf 0}(\beta K, {\bf l}) 
= \frac12 {\bf l}K{\bf l}^{\bot}+ {\bf l}K\beta^{\bot}= 
-\frac12 {\bf l}K({\mathbb I}-{\bf l})^{\bot}=
-\frac12 \sum_{i\neq j} \ve_{ij} K_{ij},$$
where $\ve_{ij}=1$ if ``$l_i=1$ and $l_j=0$'' and otherwise $\ve_{ij}=0$.
Since $\ve_{i,i\pm1}$ is not zero by the assumption, 
$f_{\bf 0}(\beta K, {\bf l})$ is positive. 
For ${\bf l}\in \{0,-1\}^g \setminus \{{\bf 0}\}$, we have  
$$f_{\bf 0}(\beta K, {\bf l}) 
= \frac12 {\bf l}K{\bf l}^{\bot}-\frac12 {\bf l}K{\mathbb I}^{\bot}\geq 
\frac12 {\bf l}K{\bf l}^{\bot} > 0.$$
Thus we see $f_{\bf 0}(\beta K, {\bf l}) > 0$ for 
${\bf l} \in \{0,1\}^g \cup \{0,-1\}^g \setminus \{ {\bf 0}, \mathbb{I} \}$
and $f_{\bf 0}(\beta K, \mathbb{I}) = 0$.
This implies $\beta K \in \partial D_{\bf 0}$
from Lemma \ref{lemma:Dm-domain}.
\\
(ii) 
Set ${\bf Z}=\beta K + nL{\bf e}_1+t\vec{\lambda}$ 
($n=0,1$, $t=0,1,2$). For ${\bf l}={\mathbb I}$,
we have 
$$f_{\bf 0}({\bf Z}, \mathbb{I}) = 
\frac12 {\mathbb I}K{\mathbb I}^{\bot}+{\mathbb I}
(\beta K+ nL{\bf e}_1+t\vec{\lambda})^{\bot}
= {\mathbb I}(nL{\bf e}_1+t\vec{\lambda})\geq 0,\quad (\mbox{by (i)}),$$
where the equality holds iff $n=t=0$.
For ${\bf l}\in\{0,1\}^g \setminus \{{\bf 0}, {\mathbb I}\}$, we have
$$f_{\bf 0}({\bf Z}, {\bf l}) = 
\frac12 {\bf l}K{\bf l}^{\bot}
+{\bf l}(\beta K+ nL{\bf e}_1+t\vec{\lambda})^{\bot}
>  {\bf l}(nL{\bf e}_1+t\vec{\lambda})\geq 0,\quad (\mbox{by (i)}).$$
From Lemma~\ref{K-etc}, we have
\begin{align*}
\beta K+ nL{\bf e}_1+t\vec{\lambda}
=& -\frac12 p_g {\bf e}_g+(n-\frac12) L {\bf e}_1+(t-1)\vec{\lambda}\\
=& -\beta K-p_g {\bf e}_g - (1-n)L {\bf e}_1-(2-t)\vec{\lambda}.
\end{align*}
Thus, for ${\bf l}\in\{0,-1\}^g \setminus \{{\bf 0}\}$, 
$f_{\bf 0}({\bf Z}, {\bf l})$ becomes
\begin{align*}
\frac12 {\bf l}K{\bf l}^{\bot}
&+{\bf l}(-\beta K -p_g {\bf e}_g- 
(1-n)L{\bf e}_1-(2-t)\vec{\lambda})^{\bot}\\
>&  -{\bf l}(p_g {\bf e}_g+(1-n)L{\bf e}_1+(2-t)\vec{\lambda})\\ 
\geq& 0 \quad (n=0,1,\ t=0,1,2). 
\end{align*}
Then the claim follows from Lemma \ref{lemma:Dm-domain}.
\\
(iii) When $\beta K+{\bf v}$ is in the interior of $D_{{\bf 0}}$,
$f_{\bf 0}(\beta K+{\bf v}, {\bf l}) 
= \frac12 {\bf l}K{\bf l}^{\bot}+ {\bf l}(\beta K+{\bf v})^{\bot}> 0$ holds
for all ${\bf l} \in \Z^g$.
Thus we see that $f_{\mathbb{I}}(\beta K-{\bf v}, -{\bf l}) > 0$
for all ${\bf l} \in \Z^g$, since 
$$f_{\mathbb{I}}(\beta K-{\bf v}, -{\bf l}) =
-{\bf l}K({\mathbb I}-\frac12 {\bf l})^{\bot}-{\bf l}(\beta K-{\bf v})^{\bot}= 
f_{\bf 0}(\beta K+{\bf v}, {\bf l}),$$
for all ${\bf l} \in \Z^g$.
\end{proof}

\begin{example}[Counter example]\label{ex-counter}
Set as $g=2$, $C_2=0$, $L=11$, $\lambda_1=2$, $\lambda_2=3$, 
$\beta = (-\frac12, 0)$
and take $P_1,P_2,P_3,P_4$ as Proposition~\ref{bilinear-theta}, then
\begin{align*} &K=\left(
\begin{array}{cc}
18& -3\\ -3& 6
\end{array}\right), \quad 
P_1=(3, 6), 
P_2=(0,0),\
P_3=(0,11),\ 
P_4=(3, 5),
\end{align*}
\begin{align*}
&\int_{P_1}^{P_3}=\int_{P_2}^{P_4}=(2,1), \quad
\int_{P_2}^{P_3}=(-11,0), \quad
\int_{P_1}^{P_4}=(15, 2), \quad
\int_{P_3}^{P_4}=\int_{P_1}^{P_2}=(13,1), 
\end{align*}
hold, and thus $\beta K + \int_{P_1}^{P_4}=(6,3.5)$ is on the boundary 
$D_{(0,-1)} \cap D_{(-1,-1)}$.
Therefore we cannot determine the sign $s_1$, while at ${\bf Z}=(0,0)$
we obtain that 
\begin{align*}
F_1=&\Theta(2,1)+\Theta(2,1)
 + \beta K \beta^{\bot} +\beta (11,0)^{\bot} +\Theta(2,1.5)
+\beta(15,2)^{\bot}+ \Theta(6,3.5)
= -9\\
F_2=&
\Theta(-11,0)+\Theta(15,2)
 +\beta K \beta^{\bot} +2\beta (-2,-1)^{\bot} +2\Theta(-11,0.5)
=-7.5\\
F_3=&
\Theta(4,2)+\Theta(0,0)
 + \beta K \beta^{\bot} +\beta(13,1)^{\bot} +\Theta(4,2.5)
+ \beta(13,1)^{\bot}+ \Theta(4,2.5)
=-8.5, 
\end{align*}
and thus Fay's type identities do not hold in this case. 
\end{example}

From Proposition~\ref{bilinear-theta} we have the following. 
\begin{corollary}\label{prop:tau-theta}
For ${\bf Z}_0 \in \R^g$, the function $T_n^t$ given by
\begin{align}
  \label{tau-theta}
  &T_n^t =   \Theta({\bf Z}_0- n L {\bf e}_1+\vec{\lambda} t )
\end{align}
satisfies the tropical bilinear equation \eqref{UD-tau}.
\end{corollary}

\medskip

%%%%%%%%%%%%%%%%%%%%%%%%%%%%%%%
\section{Solution of UD-pToda}
%%%%%%%%%%%%%%%%%%%%%%%%%%%%%%%

%%%%%%%%%%%%%%%%%%%%%%%%%%%%%%%%%%%%%%%%%
\subsection{$\tau$-function for UD-pToda}
%%%%%%%%%%%%%%%%%%%%%%%%%%%%%%%%%%%%%%%%%

Fix a positive integer $g$.
The UD-pToda is defined 
by the piecewise-linear map
$$\varphi_{\mathcal{T}}: (Q_n^t,W_n^t)_{n=1,\cdots,g+1} \mapsto 
     (Q_n^{t+1},W_n^{t+1})_{n=1,\cdots,g+1}$$  
given by \eqref{UD-pToda}
on the phase space $\mathcal{T}$ \eqref{phasespace-T}.
The map $\varphi_{\mathcal{T}}$ preserves the tropical 
polynomials $C_i(Q,W) ~(i=-1,0,\cdots,g)$
on $\mathcal{T}$. 
Fix $C = (C_{-1},C_0,\cdots,C_g) \in \R^{g+2}$ as 
\eqref{CD-condition}, and
define the tropical curve $\tilde{\Gamma}$ \eqref{trop-curve} and 
the isolevel set $\mathcal{T}_C$ as
\begin{align}
  \mathcal{T}_C = \{ (Q,W) \in \mathcal{T} ~|~ C_i(Q,W) = C_i 
                     ~(i=-1,0,\cdots,g)\}.
\end{align}
See \cite{InoueTakenawa07} for a detail of $C_i(Q,W)$.

%%%%%%%%%%%%%%%%%%%%%%%%%%%%%%%%%%%%

Let $\cS_t$ ($t\in \Z$) be a subset of infinite dimensional 
space:
$$
  \cS_t = \{T_n^t \in \R ~|~ n \in \Z \},
$$
where $T_n^t$ has a quasi-periodicity; i.e. $T_n^t$ satisfies 
$T_{n+g+1}^t = T_n^t + c_n^t$, where $c_n^t$ satisfies 
\begin{align}\label{c-period}
   ~c_n^t =an+bt+c  
\end{align}
for some $a,b,c \in \R$.
Fix $L \in \R$ such that 
\begin{align}\label{c-period2}
2b-a< (g+1)L
\end{align} and define 
a map $\varphi_{\cS}$ from $\cS_{t} \times \cS_{t+1}$ 
to $\cS_{t+1} \times \cS_{t+2}$ as
$\varphi_{\cS}: (T_n^t,T_n^{t+1})_{n \in \Z} \mapsto
 (T_n^{t+1},T_n^{t+2})_{n \in \Z}$
with 
\begin{align} \label{UD-ptau}
  &T_n^{t+2}=2 T_n^{t+1} -T_n^t + X_{n+1,t}^{(g)},
\end{align}
where we define a function on $\cS_{t} \times \cS_{t+1}$:
\begin{align}\label{X-tau} 
  X_{n,t}^{(k)} 
  = \min_{j=0,\cdots,k}\bigl[j L + 2 T_{n-j-1}^{t+1} + T_n^t+T_{n-1}^t
          - (2 T_{n-1}^{t+1} + T_{n-j}^t + T_{n-j-1}^t)\bigr],
\end{align}
for $k \in \Z_{\geq 0}$.
Note that it follows from \eqref{c-period} and \eqref{c-period2}
that $2 c_n^{t+1} - c_n^t - c_{n+1}^t < (g+1) L$ for all $n \in \Z$.
The function $X_{n,t}^{(k)}$ has the following properties:
\begin{lemma}\label{lemma:X}
  {\rm (i)} 
  $X_{n,t}^{(k)}$ satisfies a recursion relation:
  \begin{align}\label{tau-rec}
  2 T_{n-1}^{t+1} + X_{n,t}^{(k)} 
  = \min \bigl[ 2 T_{n-1}^{t+1}, L + 2 T_{n-2}^{t+1} + T_n^t
                -T_{n-2}^t+X_{n-1,t}^{(k-1)}],
  \text{ for $k \geq 1$.}
  \end{align}
  {\rm (ii)} $X_{n,t}^{(g+1)} = X_{n,t}^{(g)}$.
\end{lemma}
\begin{proof}
(i) It is shown just by the definition of $X_{n,t}^{(k)}$.
\\
(ii) 
For simplicity we rewrite \eqref{X-tau} as 
$X_{n,t}^{(k)} = \min_{j=0,\cdots,k}[a_j^t]$ where
$$
  a_j^t = j L + 2 T_{n-j-1}^{t+1} + T_n^t+T_{n-1}^t
          - (2 T_{n-1}^{t+1} + T_{n-j}^t + T_{n-j-1}^t).
$$
By making use of \eqref{c-period} we obtain
\begin{align*} 
  a_{g+1}^t &= (g+1) L + 2 T_{n-g-2}^{t+1} + T_n^t+T_{n-1}^t
             - (2 T_{n-1}^{t+1} + T_{n-g-1}^t + T_{n-g-2}^t)
            \\        
            &= (g+1) L + c_{n-g-2}^t +c_{n-g-1}^t - 2 c_{n-g-2}^{t+1} > 0.
\end{align*}
At the same time we have $a_0^t = 0$.
Thus we obtain
$$
  X_{n,t}^{(g+1)} 
  = \min[X_{n,t}^{(g)}, a_{g+1}^t] 
  = X_{n,t}^{(g)}.
$$
\end{proof}

Let $\sigma_t$ be a map $\sigma_t : 
\cS_{t} \times \cS_{t+1} \to \mathcal{T}$ 
given by
\eqref{WQ-T}. 
It is easy to check that $\sigma_t$ is well-defined
where \eqref{c-period} assures the periodicity of $(Q_n^t, W_n^t)_n$,
and \eqref{c-period2} assures the condition 
$\sum_n^{g+1} Q_n^t < \sum_n^{g+1} W_n^t$ of 
the phase space $\mathcal{T}$.

\begin{lemma}\label{lemma:tau-QW}
  The relation $\sigma_t(X^{(g)}_{n,t}) = X_n^t$ holds, and 
  the following diagram is commutative:
  \begin{align}\label{sigma-T}
    \begin{matrix}
    \cS_{t} \times \cS_{t+1} & \stackrel{\sigma_t}{\to} &  \mathcal{T} \\[1mm] 
    \quad \downarrow_{\varphi_{\cS}} & &
    \quad \downarrow_{\varphi_{\mathcal{T}}} \\[1mm]
    \cS_{t+1} \times \cS_{t+2} & \stackrel{\sigma_{t+1}}{\to} & \mathcal{T}
    \end{matrix}~~.
  \end{align}
\end{lemma}
\begin{proof}
  By direct calculation we can check $\sigma_t(X^{(g)}_{n,t}) = X_n^t$.
  To check $\varphi_{\mathcal{T}} \circ \sigma_t = \sigma_{t+1} \circ 
\varphi_{\cS}$,
  it is enough to calculate $Q_n^{t+1}$ in the image of each map. 
  We have
  \begin{align*}
    Q_n^{t+1} 
    &= \min[W_n^t, Q_n^t - X_n^t] \\
    &= \min[L + T_{n-1}^{t+1} + T_{n+1}^{t} - T_{n}^{t} - T_{n}^{t+1},
            T_{n-1}^{t} + T_{n}^{t+1} - T_{n-1}^{t+1} - T_{n}^{t} 
           - X_{n,t}^{(g)}]  ~~ \text{ (by \eqref{WQ-T})},
  \end{align*}
  for $\varphi_{\mathcal{T}} \circ \sigma_t$, and for $\sigma_{t+1} \circ 
\varphi_{\cS}$,
  \begin{align*}
    Q_n^{t+1}
    &= T_{n}^{t+2} + T_{n-1}^{t+1} - T_{n-1}^{t+2} - T_{n}^{t+1} \\
    &= 2 T_n^{t+1} + X_{n+1,t}^{(g)}
       + (T_{n-1}^{t+1} - T_n^t  - T_{n-1}^{t+2} - T_{n}^{t+1})
       ~~~ \text{ (by \eqref{UD-ptau})} \\
    &= 2 T_n^{t+1} + X_{n+1,t}^{(g+1)}      
    + \underline{(-X_{n,t}^{(g)}-T_{n-1}^{t+1} -T_n^t - T_n^{t+1}+T_{n-1}^t)}
     ~~~ \text{ (by Lemma \ref{lemma:X}(ii) and \eqref{UD-ptau})} \\
    &= \min[2 T_n^{t+1}, L + 2 T_{n-1}^{t+1} + T_{n+1}^t
                -T_{n-1}^t+X_{n,t}^{(g)}] + \underline{( \qquad )} 
       ~~~ \text{ (by \eqref{tau-rec})}.
  \end{align*}
  It is easy to see these two expressions of $Q_n^{t+1}$ coincide.  
\end{proof}

From Lemma~\ref{lemma:X}, the next proposition immediately follows.
\begin{proposition}
 If $\{T_n^t\}_{n,t}\in \prod_t \cS_t $ satisfies \eqref{UD-ptau},
 then it satisfies \eqref{UD-tau}. 
\end{proposition}

\begin{proof}
\begin{align*}
\min[2 \, T_{n}^{t+1}, ~T_{n-1}^{t+2} + T_{n+1}^{t} +L]
=& \min[2 \, T_{n}^{t+1}, 
(2 T_{n-1}^{t+1}-T_{n-1}^t + X_{n,t}^{(g)}) 
+ T_{n+1}^{t} +L]\\
=&2 T_n^{t+1} + X_{n+1,t}^{(g+1)}=2 T_n^{t+1} + X_{n+1,t}^{(g)}\\
=& T_{n}^{t} +  T_{n}^{t+2}
\end{align*}
\end{proof}

Conversely the following proposition holds,
which proved after Theorem~\ref{theta-tau} is proved.
\begin{proposition}\label{prop:tau-ptau}
  Let $\{T_n^t\}_{n,t}\in \prod_t \cS_t$ satisfy \eqref{UD-tau} and
  ${\mathcal A}_t$ denote a set:  
$$\{n \in \Z ~|~ T_{n}^{t} +  T_{n}^{t+2} = 2 T_{n}^{t+1},
  \mbox{ i.e. } 2 T_{n}^{t+1} \leq T_{n-1}^{t+2} + T_{n+1}^{t} + L\}.$$ 
  Then we have the followings:
  \begin{align*}
  {\rm (i)} ~n \in {\mathcal A}_t ~\Leftrightarrow~ n+g+1 \in {\mathcal A}_t,
  \quad
  {\rm (ii)} ~{\mathcal A}_t \neq \emptyset,
  \quad
  {\rm (iii)} ~\text{\eqref{UD-ptau} holds for all } n,t \in \Z^2.
  \end{align*}
\end{proposition}

The following is the main result of this section.
\begin{theorem}\label{theta-tau}
  Let $\iota_t : \R^g \to \cS_t \times \cS_{t+1}$ be a map:
  $$ 
    {\bf Z}_0 \mapsto 
    (T_n^t= \Theta({\bf Z}_0- n L {\bf e}_1+t \vec{\lambda}),~
   T_n^{t+1}=\Theta({\bf Z}_0- n L {\bf e}_1+(t+1)\vec{\lambda}))_{n \in \Z}.
  $$
  Then the following diagram is commutative:
  \begin{align}\label{iota-T}
    \begin{matrix}
    \R^g & \stackrel{\iota_t}{\to} & \cS_t \times \cS_{t+1}
    & \stackrel{\sigma_t}{\to} &  \mathcal{T}_C 
    \\[1mm] 
    ~\downarrow_{{\rm id.}} & & 
     \quad \downarrow_{\varphi_{\cS}} 
    & & \quad \downarrow_{\varphi_{\mathcal{T}}} \\[1mm]
    \R^g & \stackrel{\iota_{t+1}}{\to} & \cS_{t+1} \times \cS_{t+2} & 
    \stackrel{\sigma_{t+1}}{\to} & \mathcal{T}_C
    \end{matrix}\quad .~~
  \end{align}
\end{theorem} 

In short,
for any ${\bf Z}_0\in \R^g$, 
$T_n^t=\Theta({\bf Z}_0-nL{\bf e}_1+t\vec{\lambda})$ 
satisfies \eqref{UD-ptau} and gives a solution of \eqref{UD-pToda} 
through \eqref{WQ-T}.

\begin{proof}[Proof of Theorem \ref{theta-tau}]
  We first check that 
  $T_n^t=\Theta({\bf Z}_0- n L {\bf e}_1+t \vec{\lambda})$ 
  satisfies the quasi-periodicity \eqref{c-period} and \eqref{c-period2}.
  From Prop.~\ref{Theta-period} and Lemma~\ref{K-etc} (iv),
  we have
  \begin{align}\label{T-qperiod}
  T_{n+g+1}^t = T_n^t + c_n^t, 
  \qquad c_n^t =an+bt+c= \vec{g} \cdot 
  ({\bf Z}_0-nL\vec{e_1}+t \vec{\lambda} - \frac{1}{2} \vec{g} K)^t.
  \end{align}
  Then from \eqref{g-lambda} we obtain 
  \begin{align*}
    &a=-gL,\ b=2\sum_{i=1}^g (\lambda_i-\lambda_0),
    \ c=\vec{g}\cdot {\bf Z}_0-\frac12 g(g+1)L,
    \\
    &(g+1)L-2b+a= (g+1)L-2\sum_{i=1}^g (\lambda_i-\lambda_0)+gL=p_g > 0.
  \end{align*}

  Due to Prop. \ref{prop:tau-ptau} (iii),
  the left part of the diagram \eqref{iota-T} 
  is commutative.
  Since $\Theta({\bf Z})$ is associated to $\Gamma$,
  the image of $\sigma_t$ is in $\mathcal{T}_C$, while
  the commutativity of the right part of \eqref{iota-T}  follows 
  from Lemma \ref{lemma:tau-QW}.
\end{proof}
  
\begin{remark}
  f ${\mathcal A}_t= \emptyset$,
  the map "$\sigma_{t+1}\circ \varphi_{\cS} \circ \sigma_t^{-1}$" 
  induces a map given by $
  (Q_n^t,W_n^t)_{i=1,\cdots,g+1} \mapsto 
     (Q_n^{t+1}=W_n^t,W_n^{t+1}=Q_{n+1}^t)_{n=1,\cdots,g+1}$,
which does not preserve the inequality $\sum_n Q_n^t < \sum_n W_n^t$.
\end{remark}

\begin{proof}[Proof of Proposition \ref{prop:tau-ptau}]
  \begin{align*}
  \mathrm{(i)}~
    n \in {\mathcal A}_t 
    \Leftrightarrow& 
    T_{n}^{t} +  T_{n}^{t+2} = T_{n-1}^{t+2} + T_{n+1}^{t} + L
    \\
    \Leftrightarrow& 
    (T_{n+g+1}^{t} - c_n^t) + (T_{n+g+1}^{t+2} - c_n^{t+2}) 
    = (T_{n+g}^{t+2} - c_{n-1}^{t+2}) + (T_{n+g+2}^{t}-c_{n+1}^{t}) + L\qquad
    \\
    \Leftrightarrow&
    T_{n+g+1}^{t} + T_{n+g+1}^{t+2} = T_{n+g}^{t+2} + T_{n+g+2}^{t} + L
    \\
    \Leftrightarrow&     
    n+g+1 \in {\mathcal A}_t.
  \end{align*}
(ii)
Assume that $T_n^t$ satisfies the quasi-periodicity \eqref{c-period}
and ${\mathcal A}_t =\emptyset$.
  Then we have 
  \begin{align*}
    &T_n^t + T_n^{t+2} = T_{n-1}^{t+2} + T_{n+1}^t + L 
    ~~\text{ for all $n$},
    \\
    \Rightarrow& T_{n+g}^{t+2}-T_{n-1}^{t+2}= T_{n+g+1}^t-T_n^t+ (g+1)L
    \\
    \Rightarrow& -a+2b+c=c+(g+1)L
    \\
    \Leftrightarrow& 2b-a = (g+1)L.
  \end{align*}
  This contradicts the claim of the quasi-periodicity.
\\
(iii)
From (i) and (ii), it is enough to show that 
if $n \in {\mathcal A}_t$ then 
$T_{n+s}^{t+2}=2T_{n+s}^{t+1}-T_{n+s}^t+X_{n+s+1,t}^{(g)}$ holds for all 
$s\geq 0$.
We show it by induction on $s$. \\ 
In case $s=0$.
Due to \eqref{UD-tau},
\begin{align}
T_{n-j}^{t+2} &\leq 2 T_{n-j}^{t+1}-T_{n-j}^t \label{ichi}\\
T_{n-j}^{t+2} &\leq L+T_{n-j-1}^{t+2}+ T_{n-j+1}^{t}-T_{n-j}^t \label{ni}
\end{align}
hold for all $j$.
Taking the sum of \eqref{ni} for 
$j=0,1,\dots,k-1$ and using \eqref{ichi}, 
we have
\begin{align*}
T_{n}^{t+2} &\leq  kL+T_{n-k}^{t+2}+ T_{n+1}^{t}-T_{n-k+1}^t \\
&\leq kL+(2T_{n-k}^{t+1}-T_{n-k}^t)+ T_{n+1}^{t}-T_{n-k+1}^t \\
&=2T_n^{t+1}-T_n^t+a_k^t,
\end{align*}
for all $k\geq 0$,
where $X_{n+1,t}^{(g)}= \min_{k=0,1,\dots,g}[a_k^t]$,
and thus, $T_{n}^{t+2}\leq 2T_{n}^{t+1}-T_{n}^t+X_{n+1,t}^{(g)}$
holds. Since $T_{n}^{t+2}= 2T_{n}^{t+1}-T_{n}^t$ from the assumption 
$n \in {\mathcal A}_t$
and $X_{n+1,t}^{(g)}\leq a_0^t=0$, we have $X_{n+1,t}^{(g)}=0$, 
which yields the claim for $s=0$.

Assume $T_{n+s}^{t+2}=2T_{n+s}^{t+1}-T_{n+s}^t+X_{n+s+1,t}^{(g)}$ holds for 
some $s\geq 0$.
\begin{align*}
T_{n+s+1}^{t+2}=&2T_{n+s+1}^{t+1}-T_{n+s+1}^t  
+\min[0,T_{n+s}^{t+2}+T_{n+s+2}^t+L-2T_{n+s+1}^{t+1}] \quad
\mbox{(by \eqref{UD-tau})} \\
=&2T_{n+s+1}^{t+1}-T_{n+s+1}^t
+\min[0, (2T_{n+s}^{t+1}-T_{n+s}^t+X_{n+s+1,t}^{(g)})
+T_{n+s+2}^t+L-2T_{n+s+1}^{t+1}]\\
=&2T_{n+s+1}^{t+1}-T_{n+s+1}^t+X_{n+s+2,t}^{(g+1)}\quad 
\mbox{(by Lemma~\ref{lemma:X})}\\
=&2T_{n+s+1}^{t+1}-T_{n+s+1}^t+X_{n+s+2,t}^{(g)}.
\end{align*}
Thus, we have the claim holds for $s+1$. \end{proof}

\begin{corollary}\label{J-Toda}
  We write $\varphi_{\vec{\lambda}}$ 
  for the translation on the Jacobian $J(\Gamma)$ 
  as $\varphi_{\vec{\lambda}} : [{\bf Z}_0] \mapsto [{\bf Z}_0 + 
\vec{\lambda}]$.
  Let $\iota_\sigma : J(\Gamma) \to \mathcal{T}_C$ 
  be the map induced by $\sigma_t \circ \iota_t : \R^g \to \mathcal{T}$. 
  Then the following diagram is commutative:
  \begin{align*}
    \begin{matrix}
    J(\Gamma) & \stackrel{\iota_{\sigma}}{\to} &  \mathcal{T}_C \\[1mm] 
    ~\downarrow_{\varphi_{\vec{\lambda}}} & & \downarrow_{\varphi_{\mathcal{T}}} \\[1mm]
    J(\Gamma) & \stackrel{\iota_{\sigma}}{\to} & \mathcal{T}_C
    \end{matrix}~~.
  \end{align*}
\end{corollary}

\begin{proof}
The commutativity follows from \eqref{sigma-T} and \eqref{iota-T}.
The well-definedness, 
$\iota_\sigma ({\bf Z}_0) = \iota_\sigma ({\bf Z}_0+ {\bf l} K)$ 
for any ${\bf l} \in \Z^g$, is guaranteed by the quasi-periodicity 
\eqref{c-period} $T_n^t=\Theta({\bf Z}_0- n L {\bf e}_1+\vec{\lambda} t)$ 
satisfies. 
\end{proof}

Further, by concrete computation, 
we conjecture that the map $\iota_\sigma$ is isomorphic
\footnote{ 
We proved this conjecture after submitting the present paper
\cite{InoueTakenawa09}.
The proof needs rather complicated combinatorial discussion.}.

%%%%%%%%%%%%%%%%%%%%%%%%%%%%%%%%%%%%%%%%%%%%%%%%%%%%%%%%%%%%%%%%%

\appendix

%%%%%%%%%%%%%%%%%%%%%%%%%%%%%%%%%%%%%%%%%%%%%%%%
\section{Proof of Theorem~\ref{UDAJ}
(UD-limit of Abel-Jacobi map)}
%%%%%%%%%%%%%%%%%%%%%%%%%%%%%%%%%%%%%%%%%%%%%%%%

We assume $(c_{-1},c_0,\cdots,c_g) \in \R^{g+2}$.
By setting
\begin{align}\label{uv-xy}
  \begin{split}
  u&:=-x, \\
  \Delta(u)&:= (-1)^{g+1}u^{g+1}+(-1)^gc_{g} u^{g}+
  \cdots+(-1)c_1 u+c_0,\\
  v&:=2y+\Delta(u),
  \end{split}
\end{align}
the spectral curve $\gamma$ \eqref{complex-curve} becomes
the standard form of a hyperelliptic curve:
$$\gamma^1:\quad v^2=\Delta(u)^2-4c_{-1}.$$
It is known that $\Delta(u)$ has special properties
\cite{TanakaDate76,Toda81}:\\
(i) the zeros of $\Delta(u) = \prod_{i=0}^g (u_i-u)$ are 
  simple and positive, ordered as 
\begin{align}
  &0<u_0<u_1<\cdots <u_g.
\end{align} 
(ii) The zeros of $\Delta(u)^2 - 4 c_{-1} = 
  \prod_{i=0}^g (u-u_i^{+1})(u-u_i^{-1})$ are positive 
  and they can be ordered as 
  $0<u_0^{-1}<u_0^{+1}\leq u_1^{-1}< u_1^{+1} \leq \cdots 
  u_{g-1}^{+1} \leq u_g^{-1}<u_g^{+1}$,
  \\
(iii) We have $u_i^{-1} < u_i < u_i^{+1}$. 

Further, since $|c_{-1}|$ is very small 
when we consider the case of the UD-limit of $\gamma$,
we can assume that the zero locus of $\Delta(u)^2-4c_{-1}$ 
are simple and positive, i.e. 
\begin{align}\label{u-simple}
  &0<u_0^{-1}<u_0^{+1}<u_1^{-1}< u_1^{+1} < \cdots 
  <u_g^{-1}<u_g^{+1}
\end{align}
holds.
 
Take two-sheeted covering $u_+, u_- $of $u$ with branches 
$[u_j^{-1}, u_j^{+1}]$ $(j=0,1,2,\dots,g)$
and choose the homology basis $a_1,\dots,a_g,b_1,\dots,b_g$ 
and the basis  of 
the holomorphic differentials 
$\omega_1,\dots,\omega_g \in H^0(\gamma,\Omega^1)$
as usual as in Section~\ref{faysidentity}. 
i.e. 
$\omega_j$'s are written in the form 
$$\omega_j=\frac{w_{j,g-1} u^{g-1}+
w_{j,g-2}u^{g-2}+\cdots+w_{j,0}}{v}du \quad (w_{j,k} \in \C)$$
and satisfy $(\int_{a_i}\omega_j)_{i,j}=I$, 
$(\int_{b_i}\omega_j)_{i,j}=\Omega$.
\begin{figure}
\begin{center}
\scalebox{0.6}{\includegraphics{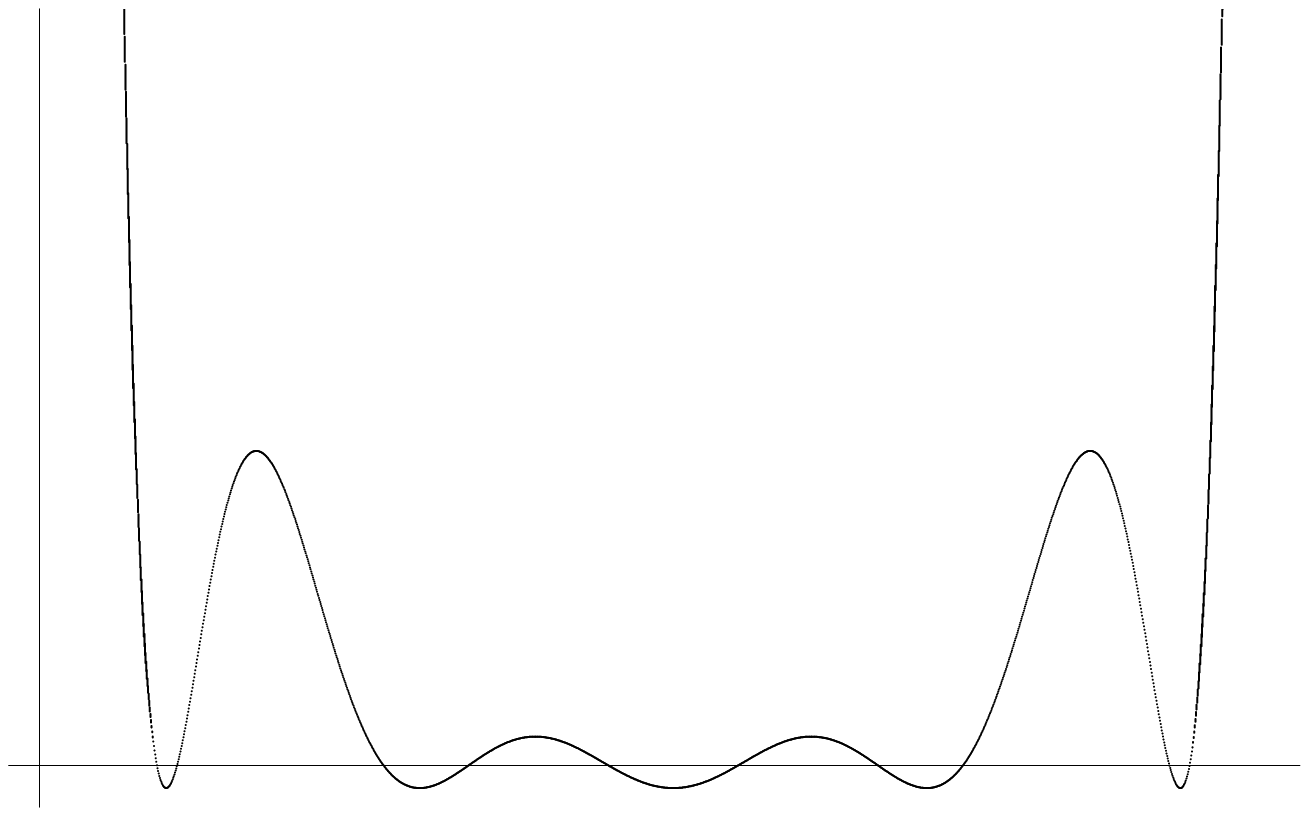}}\\
\unitlength=0.6pt
\begin{picture}(0,0)
\put(-205,18){$O$}
\put(185,18){$u$}
%\put(-215,240){$f(u)$}
\put(-170,18){{\scriptsize $u_0^{-1}$}}
\put(-140,18){{\scriptsize $u_0^{+1}$}}
\put(-100,18){{\scriptsize $u_1^{-1}$}}
\put(125,18){{\scriptsize $u_4^{-1}$}}
\put(155,18){{\scriptsize $u_4^{+1}$}}
\put(-80,0){Graph of $\Delta(u)^2-4c_{-1}$}
\end{picture}\\[2mm]
\scalebox{0.6}{\includegraphics{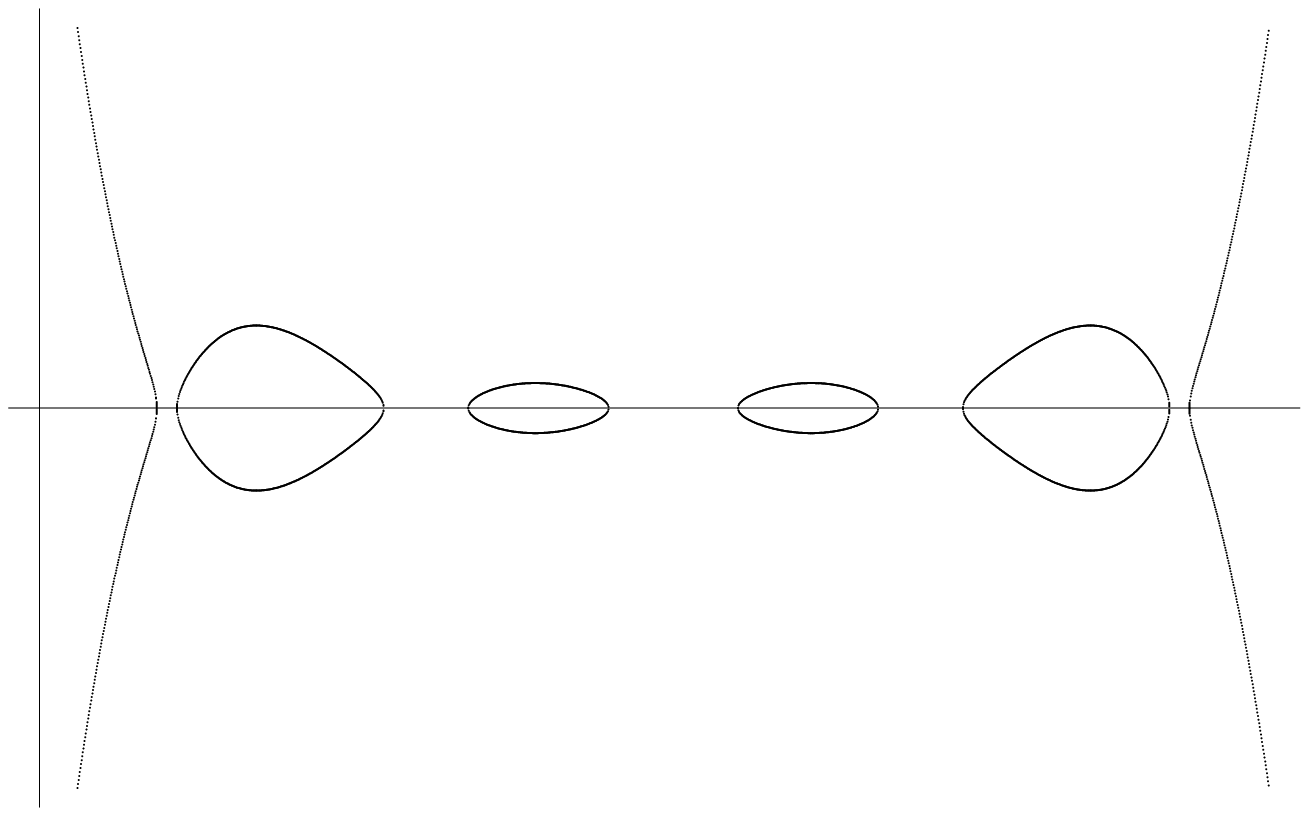}}\\
\begin{picture}(0,0)
\put(-205,125){$O$}
\put(180,125){$u$}
\put(-200,250){$v$}
\put(-80,0){$v^2=\Delta(u)^2-4c_{-1}$}
\end{picture}\\[2mm]
\scalebox{0.6}{\includegraphics{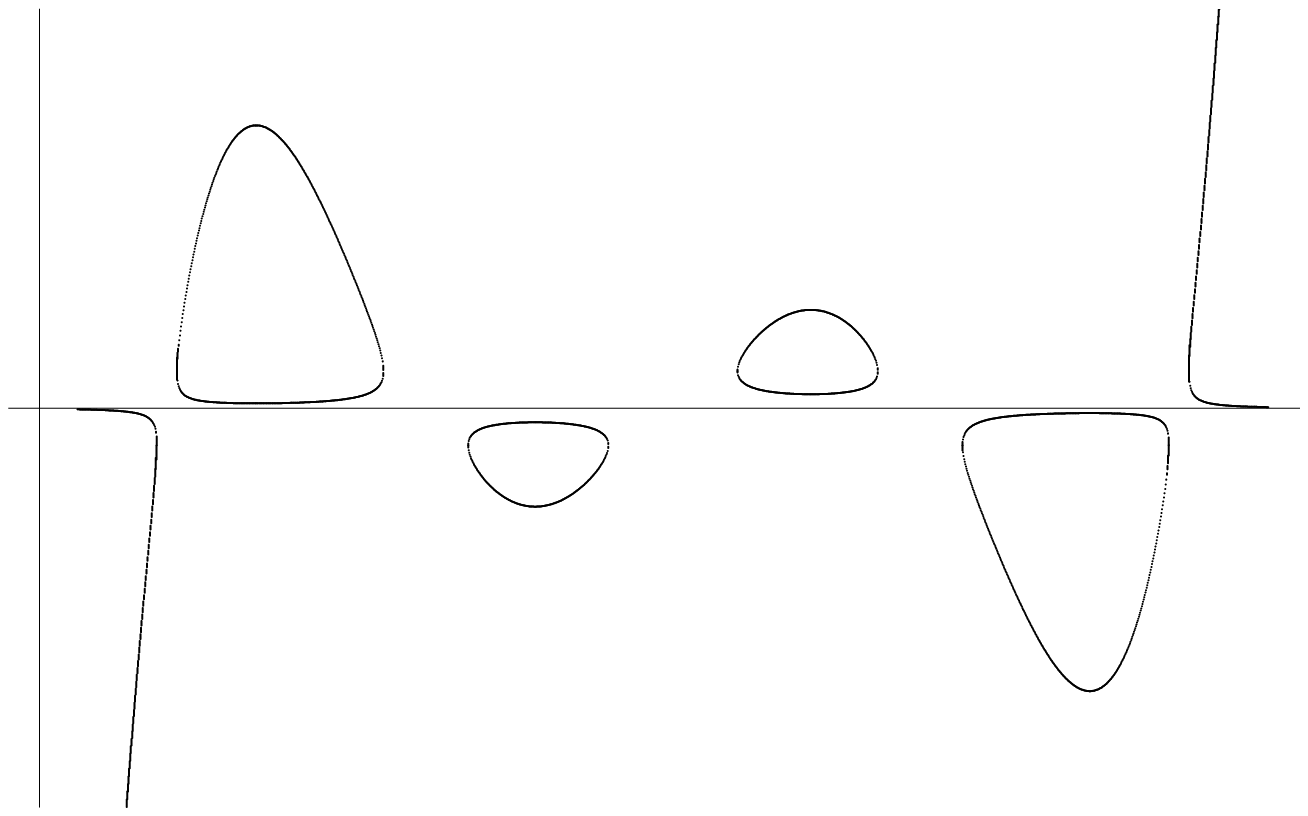}}\\
\begin{picture}(0,0)
\put(-205,125){$O$}
\put(180,125){$u$}
\put(-200,250){$y$}
\put(-80,0){$y^2+y\Delta(u)+c_{-1}=0$}
\end{picture}\\
\caption{Spectral curve with $\Delta(u)=\prod_{k=1}^5(u-(2k-1))$}
\end{center}
\end{figure}

\begin{lemma}\label{C-generic2}
When $C_i$ satisfy the generic condition \eqref{CD-condition}, 
for $0\leq i<i+2 \leq j\leq g$, 
$$C_i+C_j> C_{i+1}+C_{j-1}$$
holds.
\end{lemma}
\begin{proof}
Consider the sum of the equations $C_{i+k}+C_{i+k+2}>2C_{i+k+1}$ 
for $k=0,1,\dots,g-2$.
\end{proof}

\begin{proposition}\label{UD-XB}
By the UD-limit with the scale transformation
\begin{align} \label{scaleX} 
&|u|=e^{-\frac{X}{\ve}},\ |y|=e^{\frac{Y}{\ve}},\ 
u_j=e^{-\frac{X_j}{\ve}},\ 
u_j^{\pm 1}=e^{-\frac{X_j^{\pm 1}}{\ve}},\ 
c_i=e^{-\frac{C_i}{\ve}}, 
\end{align}
the followings holds.\\
{\rm (i)} 
$X_j=X_j^{-1}=X_j^{+1}=C_j-C_{j+1}.$
\\
{\rm (ii)} The limit of cycle $b_i$ is 
$\tilde{B}_i=B_{g-i+1}+B_{g-i+2}+\cdots+B_g$.
\end{proposition}

\begin{proof}
Notice that $X_{g}<X_{g-1}<\cdots <X_{0}$ holds by the definition of 
\eqref{scaleX}.
\\
(i) Since the even $g$ case can be shown similarly, we only show 
the odd $g$ case.
By the scale transformation, the equations $\Delta(u)=0$ and 
$\Delta(u)^2-4c_{-1}=0$
are respectively written as
$$ 
e^{-\frac{C_0}{\ve}}+e^{-\frac{2X+C_2}{\ve}}+\cdots + e^{-\frac{(g+1)X}{\ve}}
= e^{-\frac{X+C_1}{\ve}}+e^{-\frac{3X+C_3}{\ve}}+\cdots 
+ e^{-\frac{gX+C_g}{\ve}}$$
and
\begin{align*}
&\left(e^{-\frac{2C_0}{\ve}}-4e^{-\frac{C_{-1}}{\ve}}\right)+ 
\left(2e^{-\frac{2X+C_0+C_2}{\ve}}+e^{-\frac{2X+2C_1}{\ve}}\right)     
+\cdots +
 e^{-\frac{2(g+1)X}{\ve}}\\
=& 
\left(2e^{-\frac{X+C_0+C_1}{\ve}}\right)+
\left(2e^{-\frac{3X+C_0+C_3}{\ve}}+2e^{-\frac{3X+C_1+C_2}{\ve}}\right)+     
\cdots +2e^{-\frac{(2g+1)X+C_{g}}{\ve}}.
\end{align*}
By Lemma~\ref{C-generic2}, taking the UD-limit of both sides, we have
$$ 
\min[C_0, 2X+C_2,\dots, (g+1)X]=\min [X+C_1,3X+C_3,\dots,gX+C_g]$$
and
\begin{align*}
&\min[2C_0,2X+2C_1,\dots,(2g+2)X]\\
=& 
\min[X+C_0+C_1,3X+C_1+C_2,\dots,(2g+1)X+C_{g}].
\end{align*}
By solving these, we obtain the claim. \\
(ii) It is easily shown. 
\end{proof}

\begin{proposition}\label{omega}
Set $\omega_j^0$ as 
$$\omega_j^0=\frac{1}{2\pi \I}
\left\{\frac{1}{u-u_j}-\frac{1}{u-u_0}
\right\} du
$$ for $j=1,\cdots,g$ and define $u_{j,k}\in \C$ by
$$\omega_j^0=\frac{u_{j,g-1} u^{g-1}+
u_{j,g-2}u^{g-2}+\cdots+u_{j,0}}{\Delta(u)}du.$$
Set $\tilde{\omega}_j$ as
$$\tilde{\omega}_j=\frac{u_{j,g-1} u^{g-1}+
u_{j,g-2}u^{g-2}+\cdots+u_{j,0}}{
\sqrt{\Delta(u)^2-4c_{-1}}}du$$
and take the scale transformation \eqref{scaleX}, then
\begin{align} 
\lim_{\ve\to +0} \int_{a_i}\tilde{\omega}_j=
\lim_{\ve\to +0} \int_{a_i}\omega_j^0=\delta_{i,j}
\end{align}
hold.
\end{proposition}

\begin{proof}
Since $2C_0<C_{-1}$ from \eqref{CD-condition},
we have $\lim_{\ve \to +0} \tilde{\omega}_j/\omega_j^0=1$.
Further, by the residue theorem,
$\int_{a_i}\omega_j^0=\delta_{i,j}$ holds for any $\ve>0$.
\end{proof}

By this proposition, denoting the normalized 1-form on $\gamma^1$ 
as $\omega_j$, we have $\lim_{\ve \to +0} \omega_j/\omega_j^0=1$.

\begin{remark}
By the scale transformation \eqref{scaleX}, 
the integral paths $a_i$ and $b_i$
are also transformed.
Although the integral paths converge to zero 
in the variables $u (=\exp( - X/\ve))$, $u_i (=\exp( - X_i/\ve))$,
they converge to nonzero paths of finite length in the variables $X$, $X_i$.
Since we can exchange the order of the UD-limit and the integration,
we will consider by the variables $X$'s in the following.
\end{remark}

\begin{proposition}\label{Omegaij1}
Suppose that there exist the limits  
$\lim_{\ve\to +0}\Log_{\ve} (u_a,y_a)=(X_a,Y_a)$ and 
$\lim_{\ve\to +0}\Log_{\ve} (u_b,y_b)=(X_b,Y_b)$.\\
{\rm (i)}\ If $(X_a,Y_a)$ and $(X_a,Y_b)$ are on the same edge of $\Gamma$,
and $X_{i+1}<X_a,X_b <X_{i}$ for some $i$,
then
\begin{align} \label{Omegaij11}
-2 \pi \I \ve \lim_{\ve\to +0} \int_{u_a}^{u_b}\omega_j
=&\left\{
\begin{array}{ll}
0 \quad &(j\leq i )\\
(X_a-X_b) \quad &(j>i \mbox{ and } Y_a,Y_b\geq  \frac12 C_{-1})\\
-(X_a-X_b) \quad &(j>i \mbox{ and } Y_a,Y_b\leq \frac12 C_{-1})
\end{array}\right. 
\end{align}
holds. \\
{\rm (ii)}\ If $X_a=X_b=X_i$ for some $i$, then 
\begin{align} \label{Omegaij12}
-2 \pi \I \ve \lim_{\ve\to +0} \int_{y_a}^{y_b}\omega_j
=&\left\{
\begin{array}{ll}
-(Y_a-Y_b) \quad &(i=0)\\
(Y_a-Y_b) \quad &(i=j)\\
0 \quad &(i\neq 0,j)
\end{array}\right. 
\end{align}
holds.
\end{proposition}

\begin{remark}\label{branch}
The sign of $\omega_j$ is changed by passing through the branches 
$[u_k^{-1},u_k^{+1}]$. At the branch point 
$u = u_k^{\pm1}$, we have $v=0$ and $y=-\frac12 \Delta(u)=
\pm \sqrt{c_{-1}}$, where the sign is $+$ if and only if 
$u=u_k^{+1}$ ($k$ is odd) or $u=u_k^{-1}$ 
($k$ is even).
\end{remark}

\begin{proof}
For simplicity, we omit the sign of $\omega_j$ if there is no possibility of
misunderstanding. 
Since $\lim_{\ve \to 0} \omega_j/\omega_j^0=1$, the integrals are not changed 
by substituting $\omega_j$ by $\omega_j^0$. 
\\
(i) Without loss of generality we can assume $Y_a,Y_b\geq \frac12 C_{-1}$, 
which
corresponds to the sign of $\omega_j$ is $+$.
\begin{align*}
\lim_{\ve\to +0} -2 \pi \I \ve \int_{u_a}^{u_b} \omega_j^0 
=&\lim_{\ve\to +0}-2 \pi \I \ve  \int_{u_a}^{u_b}
\frac{1}{2 \pi \I}\left\{\frac{1}{u-u_j}-\frac{1}{u-u_0}
\right\} du  \\
=&-\lim_{\ve\to +0}\ve  \left[\log 
\left|\frac{(u_b-u_j)}{
(u_a-u_j)}\right|
- \log 
\left|\frac{(u_b-u_0)}{
(u_a-u_0)}\right|
 \right]\\
=&\left\{
\begin{array}{ll}
(X_b-X_a)-(X_b-X_a)=0  \quad &(j\leq i)\\
X_a-X_b  \quad &(j>i)
\end{array}\right.\\
&\quad \mbox{($\lim_{\ve\to +0} \Log_\ve(u-u')=X$ if $u>u'>0$)}.
\end{align*}
(ii) 
We divide to two cases (ii-1) 
$ Y_a, Y_b \leq \frac12 C_{-1}$ or  $ Y_a, Y_b \geq 
\frac12 C_{-1}$ and (ii-2)
$ Y_a <\frac12 C_{-1}<Y_b $ or  $ Y_a>\frac12 C_{-1}>Y_b$.
(ii-1)  Without loss of generality we can assume 
$Y_a,Y_b\geq \frac12 C_{-1}$. Substituting
$$du=-\frac{2y+\Delta(u)}{y\Delta'(u)}dy,$$
$$\Delta'(u)=\left(\prod_{k=0}^g (u-u_k)\right)'=
\Delta(u)\sum_{k=0}^g \frac{1}{u-u_k}$$
and
$$\Delta(u)=-\frac{y^2+c_{-1}}{y},$$
into $\omega_j^0$, we have
\begin{align*}
\omega_j^0 
=&\frac{1}{2\pi \I}
\left(\frac{1}{u-u_j}-\frac{1}{u-u_0} \right)
\left(\sum_{k=0}^g \frac{1}{u-u_k} \right)^{-1}
\left(-\frac{1}{y}+\frac{2y}{y^2+c_{-1}} \right) dy.
\end{align*}
Here, if $\lim_{\ve\to +0} \Log_\ve u = X_i$, 
\begin{align*}
\frac{1}{2\pi \I}
\left(\frac{1}{u-u_j}-\frac{1}{u-u_0} \right)
\left(\sum_{k=0}^g \frac{1}{u-u_k} \right)^{-1}
=&\left\{
\begin{array}{ll}
-\frac{1}{2\pi \I} \quad & (i=0)\\
\frac{1}{2\pi \I} \quad &(i=j)\\
0 \quad &(i\neq 0, j).
\end{array}\right. 
\end{align*}
Thus we have
\begin{align*}
&-2 \pi \I \ve \lim_{\ve\to +0} \int_{y_a}^{y_b}\omega_j\\
=&\left\{
\begin{array}{ll}
\displaystyle
\lim_{\ve\to +0} \ve [\log (y +\frac{c_{-1}}{y})]_{y_a}^{y_b} \quad 
& (i=0)\\
\displaystyle
\lim_{\ve\to +0}-\ve [\log (y +\frac{c_{-1}}{y})]_{y_a}^{y_b} \quad 
& (i=j)\\
\displaystyle
0 \quad &(i\neq 0, j)
\end{array}\right.\\ 
=&\left\{
\begin{array}{ll}
-\min[Y_b,C_{-1}-Y_b]+\min[Y_a,C_{-1}-Y_a]=-(Y_a-Y_b)  \quad & (i=0)\\[1mm]
\min[Y_b,C_{-1}-Y_b]-\min[Y_a,C_{-1}-Y_a]= (Y_a-Y_b) \quad & (i=j)\\[1mm]
0 \quad &(i\neq 0, j).
\end{array}\right. 
\end{align*}
(ii-2)\
Without loss of generality we can assume $Y_a <\frac12 C_{-1}<Y_b $. 
The ultra-discrete limit of $|y|=\sqrt{c_{-1}}$ is $Y = \frac12 C_{-1}$.
From (ii-1) and Remark~\ref{branch}, 
\begin{align*}
&-2 \pi \I \ve \lim_{\ve\to +0} 
\left(\int_{y_a}^{y_b}\omega_j \right)
=-2 \pi \I \ve \lim_{\ve\to +0} 
\left(\int_{y_a}^{-\frac12\Delta(u)}\omega_j
+\int_{-\frac12\Delta(u)}^{y_b}\omega_j \right)\\
=&\left\{
\begin{array}{ll}
-(\frac12C_{-1}-Y_a)+(\frac12 C_{-1}-Y_b)=Y_a-Y_b \quad &(i=0)\\[1mm]
(\frac12C_{-1}-Y_a)-(\frac12 C_{-1}-Y_b)=-Y_a+Y_b \quad &(i=j)\\[1mm]
0 \quad &(i\neq 0,j).
\end{array}\right.
\end{align*} 
\end{proof}

Theorem~\ref{UDAJ} immediately follows from Proposition~\ref{Omegaij1}
and Proposition~\ref{UD-XB} (ii).
Actually, by Proposition~\ref{Omegaij1}, Theorem~\ref{UDAJ} holds 
if the points $P$ and $Q$ are on a common edge of $\Gamma$. General
case is shown by the additivity of the Abel-Jacobi map $\eta$ for paths.
By this fact together with Proposition~\ref{UD-XB} (ii),
the UD-limit of $\Omega_{ij}$ becomes
$\tilde{K}_{ij}=\<\tilde{B}_i,\tilde{B}_j\>$.

\end{document}